\def\ba{\bar a}

\def\tt{\tilde t}

\def\be{\begin{equation}}
\def\ee{\end{equation}}
\def\te{\end{equation}}
\def\bea{\begin{eqnarray}}
\def\ba{\begin{eqnarray}}
\def\eea{\end{eqnarray}}
\def\ea{\end{eqnarray}}
\def\tea{\end{eqnarray}}
\def\nn{\nonumber}

\def\1/2{\frac{1}{2}}

\def\ba{\bar a}

\def\tt{\tilde t}

\def\be{\begin{equation}}
\def\ee{\end{equation}}
\def\te{\end{equation}}
\def\bea{\begin{eqnarray}}
\def\ba{\begin{eqnarray}}
\def\eea{\end{eqnarray}}
\def\ea{\end{eqnarray}}
\def\tea{\end{eqnarray}}
\def\nn{\nonumber}

\def\l{\lambda}

\def\(#1){(\ref{#1})}

\newskip\humongous \humongous=0pt plus 1000pt minus 1000pt

\newif\ifdtup

\documentclass[a4paper]{jpconf}
\makeatletter                    %allow @ in command names
\@addtoreset{equation}{section}  %reset equation count in each section
\makeatother                     %disallow @ in command names

\usepackage{graphicx}

\begin{document}

\title{Macroscopic Quantum Phenomena from the \\ Correlation, Coupling and Criticality Perspectives}

\author{C H Chou$^1$, B L Hu$^2$ and Y Suba\c{s}\i$^2$}
\address{$^1$Department of Physics, National Cheng Kung                                 % chou  0125
University, Tainan, Taiwan 701 and \\ National Center for                        % chou  0125
Theoretical Sciences (South), Tainan, Taiwan 701 }
\address{$^{2}$ Maryland Center for Fundamental Physics and Joint Quantum Institute,\\
University of Maryland, College Park, Maryland 20742, USA}

%\author{C. H. Chou}
%\address{Department of Physics, National Cheng Kung                                 % chou  0125
%University, Tainan, Taiwan 701} \address{National Center for                        % chou  0125
%Theoretical Sciences (South), Tainan, Taiwan 701}                                       % chou  0125
%\author{B. L. Hu and Y. Subasi}
%\address{Joint Quantum Institute and Maryland Center for Fundamental Physics, University of Maryland, College Park, Maryland 20742}

\ead{blhu@umd.edu}

\begin{abstract}

In this sequel paper we explore how  macroscopic quantum phenomena  can be measured or understood from the behavior of quantum correlations which exist in a quantum system of many particles or components and how the interaction strengths change with energy or scale, under ordinary situations and when the system is near its critical point.
%and under what conditions would quantum features persist or dominate in that limit.
We use the nPI (master) effective action related to the Boltzmann-BBGKY / Schwinger-Dyson hierarchy of equations as a tool for systemizing the contributions  of higher order correlation functions to the dynamics of lower order correlation functions. Together with the large $N$ expansion discussed in our first paper \cite{MQP1} we explore 1) the conditions whereby an H-theorem is obtained, which can be viewed as a signifier of the emergence of macroscopic behavior in the system.  We give two more examples from past work: 2) the nonequilibrium dynamics of $N$ atoms in an optical lattice under the large $\cal N$ (field components), $2PI$ and second order perturbative expansions, illustrating how $N$ and $\cal N$ enter in these  three aspects of quantum correlations, coherence and coupling strength. 3) the behavior of an interacting quantum system near its critical point, the effects of quantum and thermal fluctuations and the conditions under which the system manifests infrared dimensional reduction. %The effective IR dimension is determined by the spectrum of the Laplacian of an interacting quantum system, in particular, the lowest eigenmode if it exists.
We also discuss how the effective field theory concept bears on macroscopic quantum phenomena: the running of the coupling parameters with energy or scale imparts a dynamical-dependent  and an interaction-sensitive definition of `macroscopia'.

\end{abstract}
%\date{\small \today}
%\pacs{}
%\submitto{}
%\maketitle

%\centerline{\small {\it -- For IARD 2010 meeting, Hualien, Taiwan. Proceedings to appear in J. Physics (Conf. Series).}}

\section{Key points: Correlation, Interaction Strength, Effective Scales and Dimensions}

We continue our investigations begun in an earlier paper \cite{MQP1} (MQP1) into the  key features of
macroscopic quantum phenomena (MQP), seeking to identify and assemble the necessary ingredients toward the construction of a viable theoretical framework. Not only does this new field bear on the foundational issues of both quantum and statistical mechanics it also sees a widening range of manifestations and applications. (For background please refer to references in \cite{MQP1}). We want to examine the conditions or criteria whereby a macroscopic quantum system may take on classical attributes, and, more interestingly, that it keeps some of its quantum features. Our first paper focused on the large $N$ approximation, highlighting the subtle yet important difference between results obtained from a leading-order large-$N$ approximation, namely, the mean field theory which is quantum in nature, from a corresponding classical theory.  Here we focus on the key features of correlations and couplings in a quantum system of many particles or components, especially its behavior near the critical point, to investigate which factors contribute to, or determine a system as being macroscopic, and under what conditions would quantum features persist or dominate in that limit.

Before delving into these main subjects, it may be useful for the purpose of orientation to refresh what is already known in quantum statistical mechanics of a) how the thermodynamic properties of a quantum system depend on $N$, the number of particles present, and  b) how qualitative changes arise in a many-body quantum system, such as the formation of a Bose-Einstein condensate (BEC), one of the most commonly cited examples (the other being quantized fluxes in Josephson junctions superconductivity) of MQP.  We want to point out the qualitative differences in the meanings of `quantumness' in these familiar cases before we launch our investigations into the other quantum features we are more interested in, namely,  quantum correlations, fluctuations,  coherence and entanglement, and how they may show up or even persist at a macroscopic scale.

\subsection{$N$ dependence in quantum statistical mechanics}

Note from the start that one of the two basic premises of quantum statistical mechanics (QSM) \cite{HuangSM}, namely, the assumption of \textsl{random phase approximation}, precludes quantum coherence considerations. All descriptions are in terms of probabilities, not amplitudes. The other basic premise, that the system has \textsl{equal a priori probability } to be found in any of its accessible states, refers to the equilibrium condition. Note, however, that decoherence and quantum to classical transition \cite{ZurekPT,QdecBook} are fundamentally nonequilibrium processes.

The `quantum' in QSM refers  to effects due to \textbf{spin-statistics}, in the difference between bosons and fermions, and \textbf{distinguishability}: different combinatorics in distinguishable (classical feature) versus identical (quantum feature) particles, which underlie the Gibbs paradox \cite{Pathria}. Thus the answer to a) above is simple: the dependence of the partition function Z on $N$ is explicit, and the thermodynamic limit is well known. With  the correct counting of distinguishable versus indistinguishable particles included in the partition function, the difference between quantum and classical features in this restricted sense is also well accounted for.

QSM describes finite temperature equilibrium processes for a quantum (in the above restricted sense) system in or near equilibrium, such as in the ordinary (thermal) critical phenomena mediated by energy fluctuations in a canonical ensemble (temperature), or numbers / species fluctuations in a grand canonical ensemble (chemical potential) description, but cannot address issues related to quantum coherence and entanglement,  the central issues in quantum information. To the extent entanglement is a useful marker for quantum phase transition QSM cannot address issues of  quantum noise or fluctuation phenomena.

BEC is the well-known process where one can pinpoint the number $N_c$ of a system of $N$ bosonic atoms undergoing a transition where the salient features of the system undergo marked qualitative changes on the large scale, exemplifying MQP.  There, the critical number $N_c$ and temperature $T_c$ which define the phase transition also depend on the dimensionality. (E.g., a two-dimensional Bose gas undergoes transition into vortex-antivortex pairs -- see e.g., \cite{CHH}). BEC is certainly a quantum entity, carrying macroscopic quantum coherence described by the $N$-atom wave function as  many exquisite experiments carried out in the last decade have witnessed.   Yet its pivotal feature, its `quantumness', originates from the spin-statistics properties (bosonic vs fermionic condensate) and the critical phenomenon which describes the formation of a condensate is a classical, not a quantum phase transition. This addresses query b) above.

Making the distinction between `quantumness' arising from particle spin-statistics as in quantum statistical mechanics and that of large scale quantum coherence as in BEC offers a way to discern what would be called MQP in the modern sense of the word: usually it is the latter, not the former, which is the deciding factor. To name two examples:
1) Can the behavior of a collection of quantum gas be regarded as MQP? In solid state physics one treats the conduction electrons in a metal as a degenerate Fermi gas because the room temperature is much lower than the Fermi temperature of the metal. The electron's de Broglie thermal wavelength is much larger than the atomic spacing. But one would not regard the thermal properties of metals as MQP. Similar situations occur in the white dwarfs and neutron stars where the degenerate pressure of electrons or neutrons respectively balances the gravitation attraction to keep the stars in relative stable configurations. One does not refer to the existence of white dwarfs or neutron stars as examples of MQP.
2) Can one view collective excitations as MQP? Phonons are the quantized long wavelength modes of lattice vibrations. Like waves all such collective variables involve a large number of atoms and can span a large spatial extent. But one would not  call these collective phenomena as MQP.
%treatment is quantum statistical and the degrees of freedom participating in are macroscopically collective coordinate.
However for a large population of bosonic gas  accumulating in the ground state at an extremely low temperature forming a Bose-Einstein condensate(BEC),  we do regard it as a MQP. The crucial difference is the large scale quantum coherence established in such systems. Similarly, for superconductors: the Cooper pairs existing at large spatial separation (compared to the extent of electron wave functions) are quasiparticles with full quantum coherence \footnote{Another more practical angle or functional criterion towards this demarkation is to ask whether one could use such a system to perform quantum information processing, where quantum coherence and entanglement are essential. Cold atoms in an optical lattice, atom assembly, as well as superconducting flux qubits are the well-known viable examples. Note, however,  there are proposed QIP schemes such as the NMR based ones, which do not invoke quantum entanglement directly, yet could achieve speed-ups over classical computation.}

%So, what is MQP? Shall we call those macroscopic physical properties whose origin could not be answered within classical physics and needs quantum mechanics to explain its mechanism as MQP?

\subsection{Scaling and Interaction, RG running and Criticality}

In parallel to explicating the (rather restrictive) meaning of `quantumness' used widely in QSM mentioned above here we wish to explore the meaning of `macroscopia' in like spirit.  Macroscopia  in the most direct way conjures large numbers: many particles or components and/or large sizes or scales. Paper I focused on the `many' aspect, here we focus on the `scale' aspect. If one goes by the scale of a quantum system alone without other considerations such as the interactions (or coupling strength) amongst its underlying constituents then the quantum behavior of a macro object such as the Universe would be similar to that of a micro one, such as the internal degrees of freedom of an atom. In fact, both obey a harmonic oscillator equation of motion, albeit the former (for the wave function of the universe) has a negative spring constant.  The  scale is usually measured by the inverse mass of the relevant process. The physically  relevant scale (or mass) is different from the bare scale (or mass) by a ratio given by a renormalization constant which captures the relevant interactions involved.  Interaction is measured  by the coupling constants which vary (`run') with energy according to the renormalization group (RG) equations. %which determines  the relevant physical scale.
The infrared behavior of an interacting quantum system at or near its critical point such as the universality class it belongs to, is of special significance.  We will review briefly these familiar concepts below as the notion of `macroscopic' for quantum systems near criticality is determined by its infrared behavior and the interaction strength of its constituents.

\subsubsection{Scales, Interaction Strength and Effective Theories}

Let us examine the calculation of the energy levels of a hydrogen atom as an example. In a standard textbook treatment one uses the Schroedinger equation for an electron moving in the static Coulomb field of the proton. To a good approximation, the only properties of a
proton which are relevant to this problem are its mass and charge.
The knowledge of the quark structure of the proton is not
necessary to compute the energy levels of the hydrogen atom.

The required knowledge of the proton depends on how accurate one asks
for the energy levels. A more detailed calculation, taking
hyperfine splitting into account,  for instance, requires the
knowledge of the spin of the proton and the value of the magnetic
dipole moment. An even more accurate calculation requires the
knowledge of the proton charge radius and details of
the proton structure. % are needed if we want a more accurate answer for the energy levels.

The typical length scale characteristic of the hydrogen atom is
the Bohr radius $r_0=1/(m_e \alpha)$ (in units where $\hbar = c =1$ ). The typical momentum scale is $1/r_0= m_e
\alpha$, the typical energy scale characteristic of the hydrogen atom
is $\sim m_e \alpha^2$, and the typical time scale is $1/(m_e
\alpha^2)$. We can get a quantitative estimate of the errors
caused by the neglected interactions.

In an effective theory description the relevant interactions  also depend on
the question being asked. For the hydrogen atom the energy spectrum
can be computed to the accuracy $(m_e \alpha/ M_w)^2$ while
ignoring the weak interactions. But if one is interested in atomic
parity violation, the weak interactions are the leading
contributions because the electromagnetic and strong interactions
conserve parity. The effect of atomic parity violation will be
very small because the weak scale is much larger than the atomic
scale.

As one approaches shorter and shorter distances (or higher and
higher energies), one needs to invoke the relevant physical processes
at that scale. The high energy processes affecting low energy
physics appear in physical parameters measured at the low energy
regime. This is captured nicely by effective field theories, the essence of which we will discuss in a later section.

\subsubsection{RG running and Criticality}

In an interacting quantum system the variation of interaction strength with energy is described by the RG equations which are derived from the renormalization of the coupling constants (for ultraviolet divergence). Assuming that the system scales homogeneously this enables one to obtain the infrared properties of a quantum system from its ultraviolet behavior. %This is why RG theory plays such an important role in the study of critical phenomena which depends much on the infrared behavior.
At the critical point the correlation functions of such a system usually diverge: the system `feels', so to speak, all the way to infinity. Let us consider how this interesting physical situation bears on macroscopic quantum phenomena.

Critical phenomena involves many different length scales. Consider a physical system going through a continuous phase transition under some temperature change: the correlation length diverges as the
temperature approaches the critical temperature, and fluctuations at all length scales are involved. When the temperature lowers across the critical point, the system's order parameter will develop a nonzero value, where the relevant degrees of freedom act in a correlated manner. In this sense the system's behavior near its critical point indeed manifests in a macroscopic scale.

Note traditional treatment of critical phenomena is classical in that the external parameters, temperature, magnetic field, etc, are classical. But it could also be quantum in nature, the system being driven by quantum noise or vacuum fluctuations. Quantum phase transition is brought about by the change of the coupling parameters when the system
is at zero temperature. It results from the energy competition between different ground states as a function of the coupling parameter, and quantum entanglement in the system has been shown to provide a useful measure of such transitions. We hope to address this issue in a following paper when we turn our attention to the role of quantum entanglement in the description of MQP.

Naively one would expect that at the critical point as the correlation length goes to infinity, the system would not depend on the microscopic details such as the lattice spacing in the Ising model, or the interaction amongst the microscopic constituents. But it does, and that makes it more interesting. It is this particular aspect of MQP which fascinates us, namely, what and how known microscopic details determine a quantum system's macroscopic attributes, or, even more challenging, how to decipher a system's microscopic features from the macroscopic phenomena. An example is the dialectic relation between emergent and quantum gravity pertaining to the micro and macro structures of spacetime \cite{E/QG}.

On this point it is perhaps worth making the following observation: systems in a critical state can exist for a long time to certain classes of observers with specially interesting physical consequences, and many qualitative features in the ensuing phase depends on what came out of the transition.  An example is inflationary cosmology where it is believed that the universe underwent a first (old) or second (new) order transition at the GUT ($10^{14} GeV$) scale, with a duration of 68 e-folding time to be able to account for the entropy content ($10^{80}$ photons) in the present universe. During the exponential expansion the universe appears to be static for a local observer, its scale factor described by a de-Sitter -Einstein solution. In fact, in such an epoch between the metastable false vacuum and reheating to the true vacuum, one can  use the language of critical phenomena to describe its dynamics, replacing the evolution by scaling (see, e.g., \cite{RG2000}). The microscopic features of the inflaton field -- the quantum fluctuations which existed in the early universe -- are magnified during this phase transition into galaxies of today. The effect of inflation is like a giant zoom lens at work \cite{SILARG} and the universe we observe today is in this sense a truly macroscopic quantum phenomenon \cite{QUPON}.  The exponential red-shifting of outgoing waves from a black hole which accounts for the thermal feature of Hawking radiation has a similar nature, acting like a microscope in a pictorial description \cite{BHmicro}.

%why we may be inclined to see natural phenomena near criticality, or why nature preferentially appears to us in its critical state.

\subsection{Correlation in relation to fluctuations and coherence. Organization of this paper}
% via nPI effective action}

Unlike the situations of large $N$ where  an expansion in inverse $N$ yields  meaningful physical results, e.g., the leading order corresponds to a semiclassical limit, quantum correlations are less direct or transparent in their physical meanings. Correlation is related to fluctuations (as one form of fluctuation-dissipation relation indicates) and at the quantum level, quantum correlation can be a measure of quantum coherence (as the correlation history approach to quantum decoherence shows  \cite{CHcorhis}. It is also related to but distinct from quantum entanglement (see work of Cirac \cite{Cirac})
%one needs to distinguish  three distinct yet interconnected entities: correlation, fluctuations, and coherence, each entity can be either classical or quantum.
It would be very useful to identify the conditions whereby these three properties could offer a measure (no matter how tenuous) of the macroscopic features of a quantum system and how they signify the quantum (albeit residual) nature of a macroscopic system.  For this we use three examples to explore these aspects. Part I of this paper consisting of Sections 2 and 3 discusses the relation between the loop, large $N$ and nPI (n-particle irreducible) expansions, the first two signifying the appearance of quantum and macro features and how quantum correlations is placed in regard to them.   In Sec. 2 we use  the existence of a H theorem at the next-to-leading-order in the quantum mechanical $O(N)$ model to demarcate which order of large $N$ expansion will entropy generation first ensue and thus can be (approximately) regarded as acquiring some macroscopic attributes. In Sec.  3 we provide a concrete example of the above connections in the Bose-Hubbard model for $N$ atoms in an optical lattice. The nonequilibrium dynamics of this system is described by the equations of motion obtained from the closed-time-path ($CTP$) two particle irreducible ($2PI$) effective action. We compare results from  a large $\cal N$ expansion where $\cal N$ is the number of  fields (see Eq. (\ref{classAction}) below) under second order coupling with the exact (numerical) solution to see how quantum correlation plays out against $\cal N$, the number of components of the $N$ body wave function. Although  $\cal N$  or $N$ are not exactly indicators of the macroscopic, they offer a probe into the interplay between these three aspects: Correlation, fluctuations and coupling strength. Part II of this essay discusses the effective field theory (Sec. 4) and effective infrared dimensional reduction (Sec. 5). Though we didn't dwell on this point as we believe it is quite well known, we want to point out the fact that the critical behavior of interacting quantum systems can provide a sense of macroscopia quite different from the simplistic depictions (referring to the size of the system or how many components it possesses). Key to this is the fact that coupling strengths can vary (`run') with energy scale according to the RG equations.  In Sec. 4 we describe effective field theory, where this key notion well adopted in critical phenomena studies is put to good use in that the low energy phenomenology (or long wavelength behavior) is effectively independent of or largely insensitive to the high energy processes. In Sec. 5 we discuss the conditions where an interacting quantum system near the critical point may take on a lower-dimensional appearance. This effective infrared dimensional reduction is shown by an eigenvalue analysis: wherever the eigen-spectrum of the invariant operator of the order parameter field possesses a band structure with a gap separating the lowest mode or band from the higher sector, the infrared behavior of this system behaves effectively as that of  a lower dimension.  These examples serve to show the intricacy of even a simple notion such as `macroscopic' for a quantum system, that it is invariably tethered with the interaction strength, enters into the correlation lengths and undergoes qualitative changes when the system is close to the condition of criticality.
%pertaining to macroscopic quantum phenomena.
%scaling enters in capturing the large scale behavior of a system, especially near the critical point, where critical exponents determine the universality classes of different physical systems.
The characterization of quantum macroscopic phenomena by quantum entanglement is left for a later paper. \\

\noindent {\bf Part I:  Correlation hierarchy, $nPI$, $NLO$-large $N$ and H-theorem}
\vskip .5cm

\section{Correlation entropy and H theorem at $NLO$} %: nPI or master effective action}

 To prepare for our discussions on how quantum correlations play out in the quantum - macro issues we first introduce the nPI effective action with n=2 as example, which is a special case of the so-called master effective action \cite{mea} (MEA), with the specific model of a quantum mechanical $O(N)$ model (QMON) which we have used in our first paper (MQP1) to illustrate the large $N$ expansion. We then recount (see \cite{CH95,CH00}) how nPI is related to the loop expansion on the one hand, where one customarily takes as a measure of quantum features, and the large $N$ expansion on the other, where one may regard as signifying macroscopic features. (These latter two relations were expounded in MQP1.) We then discuss the implications of the existence of a H theorem proven in \cite{CH03} using the QMON model. In the next section we will use the example of the nonequilibrium dynamics of BEC atoms to illustrate how this threefold-relation amongst $1/N$, $2PI$ and expansion in the (second order) coupling strength pertain to quantum and macro issues.

%\subsection{Quantum vs Classical}

We paraphrase the results from \cite{CH03} where  a correlation
entropy for an interacting quantum field is proposed, and a proof of the existence of a
H-theorem for the quantum mechanical $O(N)$ model \cite{MACDH00,MDC01} is provided.
%, which may be regarded as a field theory in zero space dimensions.
For the former, Calzetta and Hu followed the paradigm of
Boltzmann-BBGKY \cite{balescu} and proposed a correlation entropy
(of the nth order) for an interacting quantum field \cite
{KB62,CH88} obtained by `slaving' the higher (n+1 th) order
correlation functions in the Schwinger-Dyson system of equations
(See \cite{CH95,CH00} and below). They then derived the closed time
path ($CTP$) \cite{ctp} two particle irreducible ($2PI$) \cite{2pi}
effective action ($EA$) for this model up to the next-to-leading
order ($NLO$) in $1/N$ \cite{CH03} and used the $CTP$ $2PI$ $EA$ to prove an H-theorem for the correlation entropy of a quantum mechanical $O(N)$ model at the  $NLO$ level. We begin with an introduction to the correlation hierarchy.

\subsection{Boltzmann-BBGKY Hierarchy and Schwinger-Dyson Equations}
\label{sec:BBGKY}

As is well known in nonequilibrium statistical mechanics, truncation of the BBGKY hierarchy at a finite nth order
yields a closed system \footnote{The truncation of the correlation hierarchy is not just an
arbitrary mathematical procedure, it reflects the fact that in
realistic conditions, measurement are of finite accuracy
associated with limited resolution of the instruments. Thus the
relevant physical degrees of freedom are often limited to the
lower end of the hierarchy, namely, the mean field and two-point
function.}. The system of equations describing the $n$-particle
distribution functions are time reversal invariant. When a causal
factorization condition such as Boltzmann's molecular chaos
assumption is imposed -- that is, assuming that the n+1 th
correlation function can be factorized into a product of n th
correlation functions initially but not finally (after the
collisions) -- time-irreversibility appears and an $H$-theorem
is obtained. This type of \emph{coarse graining\/} of the hierarchy
i.e., truncation plus causal factorization into a set of coupled
equations for the $n$-particle distribution functions is called
`slaving' in the language of \cite {CH95,CH00}. Slaving of the
n+1 -particle distribution function renders an otherwise closed
system (of the n+1 th order correlation functions) an effectively
open system (of nth order correlation functions) and ushers in
the appearance of dissipative dynamics \cite{CH88}. Noise and
fluctuations \cite{HuPhysica,HuBanff} should also appear, as
required by the fluctuation-dissipation relation (FDR)
\cite{fdr}, now manifesting not just for an open system near
equilibrium (as depicted in linear response theory), but for an
effectively open system \cite{CH95,CH00}.

%@@YSv5b,   BL revised v6

A familiar example is the Boltzmann equation for dilute gases. At sufficiently low density the description of a molecular gas by a truncated BBGKY hierarchy with only the one-particle and two-particle distribution functions may be justified. Because of the low density one may further assume that before the collisions the colliding partners are uncorrelated, i.e., the two-particle distribution function factorizes into a product of single-particle distribution functions. This is the physical basis for Boltzmann's imposition of the molecular chaos hypothesis. Of course the colliding partners become correlated after interaction. This assumed distinction between initial and final conditions is the origin of the macroscopic arrow of time, the appearance of dissipation and foe entropy generation in the Boltzmann paradigm.  FDR calls for a rightful place for fluctuations in the Boltzmann equation, as demonstrated
in the derivation of a stochastic Boltzmann equation \cite{KacLog,Spohn}.

There are other approximation schemes that allow for the derivation of the kinetic equations from the correlation hierarchy \cite{balescu}. To see the relation between correlation and interaction in this context of kinetic theory consider a system in which particles interact via a potential of the form $V(r_i,r_j)=\lambda v(r_i,r_j)$.
For weakly coupled systems, i.e. $\l\ll 1$, the hierarchy of
equations can be written for a certain order of $\lambda$.

To $O(\lambda)$ the equation  for the one-particle distribution
function, namely, the Vlasov equation, is closed, i.e., it does not
involve any two-particle distribution function. We emphasize that in
this case truncation alone \emph{not} followed by slaving is
sufficient to obtain a closed equation for the one-particle
distribution function. There is no collision term for short-ranged
interaction, but interactions between particles do exist represented
by an average potential of all other particles present. Vlasov
equation is useful for describing the kinetics of systems containing
many particles interacting under long range forces. Recall we have
seen it in our first paper as the equation obtained from the leading
order large N expansion in the O(N) model. Vlasov equation is
obtained as a mean field theory with reversible dynamics and no
H-theorem.

To $O(\lambda^2)$ the set of  equations for the one-particle and
two-particle distribution functions is also closed. By slaving the
two-particle distribution function to the one-particle distribution
function the Landau equation is obtained. Similar to the Boltzmann
equation, Landau equation also has a collision term accounting for
the collisions between individual particles at $O(\lambda^2)$. Landau
equation describes irreversible dynamics and its solutions satisfy
the H-theorem. What is different from the Boltzmann equation is the form of the collision integral. This difference can be traced back to the absence of a term in the equation for the two-particle distribution function at $O(\lambda^2)$ compared to the low density approximation.

The situation in quantum kinetic field theory is completely analogous. By
quantum kinetic field theory (see, e.g., \cite{KB62,CH88}), we are
referring to the hierarchy of coupled equations for the relativistic Wigner
function and its higher-correlation analogs, obtainable from the variation
of the master effective action \cite{CH95,CH00} whose variation yields the
Schwinger-Dyson equations. This is a quantum analogue of the BBGKY
hierarchy, expressed in a representation convenient for distinguishing
between microscopic (quantum field-theoretic) and macroscopic (transport and
relaxation) phenomena \footnote{%
It should be pointed out that in order to \emph{identify\/} the relativistic
Wigner function with a distribution function for quasiparticles, one must
show that the density matrix has \emph{decohered,\/} and this is neither
guaranteed nor required by the existence of a separation of macroscopic and
microscopic time scales \cite{Hab90}.}.

One may choose to work with a truncation of the hierarchy of the Wigner
function and its higher correlation analogs, or one may instead perform a
slaving of, for example, the Wigner-transformed four-point function, which
leads (within the context of perturbation theory) directly to the
relativistic Boltzmann equation \cite{CH88} and the usual $H$-theorem. The
truncation and subsequent slaving of the hierarchy within quantum kinetic
field theory can be carried out at any desired order, as dictated by the
initial conditions and relevant interactions.

\subsection{Entropy from Slaving the Higher Correlations: The $CTP$ $2PI$ $EA$}
\label{sec:entropy_slaving}

A general procedure has been presented for obtaining coupled
equations for correlation functions at any order $l$ in the
correlation hierarchy, which involves a truncation of the
\emph{master effective action\/} at a finite order in the loop
expansion \cite{CH95,CH00}. By working with an $l$ loop-order
truncation of the master effective action, one obtains a closed,
time-reversal invariant set of coupled equations for the first
$l+1$ correlation functions, $\hat{\phi},G,C_{3},\ldots
,C_{l+1}$. In general, the equation of motion for the highest
order correlation function will be linear, and thus can be
formally solved using Green's function methods. The existence of
a unique solution depends on supplying causal boundary
conditions. When the resulting solution for the highest
correlation function is  back-substituted into the evolution
equations for the other lower-order correlation functions, the
resulting dynamics is \emph{not\/} time-reversal invariant, and
generically dissipative. Thus, as was described before, with the
slaving of the higher-order (Wigner-transformed) correlation
function in quantum kinetic field theory, we have rendered a
closed system (the truncated equations for correlation functions)
into an \emph{effectively open system.\/} In addition to
dissipation, one expects that an effectively open system will
manifest noise/fluctuations, as shown in \cite{CH95,CH00} for the
case of the slaving of the four-point function to the two-point
function in the symmetry-unbroken $\lambda \Phi ^{4}$ field
theory. Thus a framework exists for exploring irreversibility and
fluctuations within the context of a unitary quantum field
theory, using the truncation and slaving of the correlation
hierarchy. The effectively open system framework is useful for
precisely those situations, where a separation of macroscopic and
microscopic time scales (which would permit an effective kinetic
theory description) does \emph{not\/} exist, such as is
encountered in the thermalization issue.

As a particular coarse graining measure the slaving of higher correlation functions to lower-order correlation functions within a particular truncation of the correlation hierarchy  has several important benefits. It can be implemented in a truly nonperturbative fashion. This necessitates a nonperturbative resummation of
daisy graphs, which can be incorporated in the truncation/slaving of the
correlation hierarchy in a natural way.

In \cite{CH00} the authors are interested in the growth of entropy due to the coarse
graining of the \emph{correlation hierarchy\/} by slaving a higher
correlation function. The simplest nonperturbative truncation of the
Schwinger-Dyson equations for the $\lambda \Phi^4$ field theory which
contains the time-dependent Hartree-Fock approximation
%\footnote{% In the reheating example, this is necessary for taking into account the
%large variance of the inhomogeneous modes of the inflaton field at the end of preheating}
is the two-loop truncation of the master effective action, in
which only the mean field $\hat{\phi}$, the two-point function $G$, the
three-point function $C_3$ are dynamical. All higher order correlation
functions obey algebraic constraints, and can thus be expressed in terms of
the three dynamical correlation functions.

While this truncation of the Schwinger-Dyson equations is well-defined and
could in principle be solved, it is disadvantageous because, as stated
above, without some coarse graining, the system will not manifest
irreversibility and will not equilibrate. Therefore we slave the three-point
function to the mean field and two-point function, and thus arrive at an
effectively open system. In principle, a systematic analysis of the
coarse-grained dynamics of the mean field and two-point function should
include stochasticity \cite{CH95,CH00}.

The $2PI$ formalism is also suitable for addressing this question because, provided an
auxiliary field is cleverly introduced, the $2PI$ $CTP$ effective action can
be found in closed form at each order in $1/N$ \cite{CH03,AABBS02}. We now
use the familiar quantum mechanical $O(N)$ model to illustrate how to derive the $2PI$ effective action and then explore the conditions for the existence of an H-theorem. This model was used in our first paper \cite{MQP1} for the discussion of large $N$ and macroscopia.

\subsection{$O(N)$ $\lambda X^{4}$ Theory}
\label{sec:CH03}

As we recall the system dynamics is described by the Hamiltonian with variables
$X_{A}$ and their conjugate momenta $P^{A}$, where $A,B$ are the
$O(N)$ group indices, with
\begin{equation}
\label{eq:Hamiltonian}
H={\frac{1}{2}} \left\{ P^{B}P^{B}+M^{2}X_{B}X_{B}+\frac{\lambda }{4N}\left(
X_{B}X_{B}\right) ^{2}\right\}
\end{equation}
The classical action

\begin{equation}
S=\int dt\;{\frac{1}{2}}\left\{ \dot{X}_{B}\dot{X}_{B}-M^{2}X_{B}X_{B}-\frac{%
\lambda }{4N}\left( X_{B}X_{B}\right) ^{2}\right\}
\end{equation}
where $M_{0}^{2}$ and $\lambda _{0}$ are the mass parameter and
coupling constant. We rescale $X_{B}\equiv\sqrt{N} x_{B}$

\begin{equation}
S=N\int dt\;{\frac{1}{2}}\left\{ \dot{x}_{B}\dot{x}_{B}-M^{2}x_{B}x_{B}-%
\frac{\lambda }{4}\left( x_{B}x_{B}\right) ^{2}\right\}
\end{equation}
Discarding a constant term, we may rewrite the classical action as

\begin{equation}
S=N\int dt\;{\frac{1}{2}} \left\{ \dot{x}_{B}\dot{x}_{B}-\left[ \frac{M^{2}}{%
\sqrt{\lambda }}+\frac{\sqrt{\lambda }}{2}x_{B}x_{B}\right] ^{2}\right\}
\end{equation}

To set up the $1/N$ resummation scheme, it is customary to
introduce the auxiliary field $\chi $, writing

\begin{equation}
S=\frac{N}{2}\int \left\{ \dot{x}_{B}\dot{x}_{B}-\left[ \frac{M^{2}}{\sqrt{%
\lambda }}+\frac{\sqrt{\lambda }}{2}x_{B}x_{B}\right] ^{2}+\left[ \frac{%
M^{2}-\chi }{\sqrt{\lambda }}+\frac{\sqrt{\lambda }}{2}x_{B}x_{B}\right]
^{2}\right\}
\end{equation}
whence

\begin{equation}
S=N\int dt\;\left\{ {\frac{1}{2}}\dot{x}_{B}\dot{x}_{B}-\chi \left[ \frac{%
M^{2}}{\lambda }+\frac{x_{B}x_{B}}{2}\right] +\frac{1}{2\lambda }\chi
^{2}\right\} .  \label{classact}
\end{equation}
{}From now on, we consider $\chi $ and $x_{B}$ as fundamental
variables on equal footing.

Because of the $O(N)$ symmetry, the symmetric point must be a solution of
the equations of motion. For simplicity, we shall assume we are within this
symmetric phase, and treat $x_{B}$ as a quantum fluctuation. We also split
the auxiliary field $\chi =\bar{\chi}+\tilde{\chi}$ into a background field $%
\bar{\chi}$ and a fluctuation field $\tilde{\chi}$. The action becomes

\begin{equation}
S=S_{0}+S_{1}+S_{2}+S_{3}
\end{equation}
$S_{0}$ is just the classical action evaluated at $x_{B}=0,$ $\chi =\bar{\chi%
}$:

\begin{equation}
S_{0}=\frac{N}{\lambda }\int dt\;\left\{ \frac{1}{2}\bar{\chi}^{2}-M^{2}\bar{%
\chi}\right\}
\end{equation}
$S_{1}$ contains terms linear in $\tilde{\chi}$ and can be set to
zero by a suitable choice of the background field $\bar{\chi}$:

\begin{equation}
S_{1}=\frac{N}{\lambda }\int dt\;\left\{ \bar{\chi}-M^{2}\right\} \tilde{\chi%
}
\end{equation}
$S_{2}$ contains the quadratic terms and yields the tree - level
inverse propagators,

\begin{equation}
S_{2}= N \int dt\;\left\{ {\frac{1}{2}}\dot{x}_{B}\dot{x}_{B}-\frac{\bar{\chi%
}}{2}x_{B}x_{B}+\frac{1}{2\lambda }\tilde{\chi}^{2}\right\}
\end{equation}
Finally $S_{3}$ contains the bare vertex

\begin{equation}
S_{3}=\left( \frac{-N}{2}\right) \int d^{d}x\;\left\{ \tilde{\chi}%
x_{B}x_{B}\right\}.
\end{equation}

To write the $2PI$ $CTP$ $EA$ we double the degrees of freedom,
incorporating a branch label $a=1,2$ (for simplicity, if not
explicitly written, we assume that the label $a$ includes
the time branch, i.e.,  $x^{Aa}\equiv x^{Aa}\left( t_{a}\right)
$). We also introduce propagators $G^{Aa,Bb}$ for the path ordered
expectation values

\begin{equation}
G^{Aa,Bb}=\left\langle x^{Aa}x^{Bb}\right\rangle  \label{eq12}
\end{equation}
and $F^{ab}$ for

\begin{equation}
F^{ab}=\left\langle \tilde{\chi}^{a}\tilde{\chi}^{b}\right\rangle
\end{equation}
Because of symmetry, it is not necessary to introduce a mixed propagator, for $%
\left\langle \tilde{\chi}^{a}x^{Bb}\right\rangle \equiv 0$. The $2PI$ $CTP$ $EA$ reads

\begin{eqnarray}
\Gamma &=&S_{0}\left[ \bar{\chi}^{1}\right] -S_{0}\left[ \bar{\chi}%
^{2}\right]  \nonumber \\
&&+\frac{1}{2}\int dudv\;\left\{ D_{Aa,Bb}(u,v)G^{Aa,Bb}(u,v)+\frac{N}{%
\lambda _{0}}c_{ab}\delta (u,v)F^{ab}\left( u,v\right) \right\}  \nonumber \\
&&-\frac{\rmi\hbar }{2}\left[ \Tr\;\ln G+\Tr\;\ln F\right] +\Gamma _{Q}
\end{eqnarray}
where, if the position variable is explicit, $c_{11}=-c_{22}=1$, $%
c_{12}=c_{21}=0$,

\begin{equation}
D_{Aa,Bb}(u,v)=N\delta _{AB}\left[ c_{ab}\partial _{x}^{2}-c_{abc}\bar{\chi}%
^{c}\right] \delta (u,v),
\end{equation}
and $c_{abc}=1$ when all entries are $1$, $c_{abc}=-1$ when all entries are $%
2$, and $c_{abc}=0$ otherwise. When we use the compressed notation, it is
understood that $c_{ab}\equiv c_{ab}\delta (t_{a},t_{b})$ and $c_{abc}\equiv
c_{abc}\delta (t_{a},t_{b})\delta (t_{a},t_{c})$. $\Gamma _{Q}$ is the sum
of all $2PI$ vacuum bubbles with cubic vertices from $S_{3}$ and propagators
$G^{Aa,Bb}$ and $F^{ab}.$ Observe that $\Gamma _{Q}$ is independent of $\bar{%
\chi}^{c}$.

Taking variations of the $2PI$ $CTP$ $EA$ and identifying $\bar{\chi}^{1}=%
\bar{\chi}^{2}=\bar{\chi}$, we find the equations of motion

\begin{equation}
\frac{N}{2}\delta _{AB}D_{ab}-\frac{\rmi\hbar }{2}\left[ G^{-1}\right] _{Aa,Bb}+%
\frac{1}{2}\Pi _{Aa,Bb}=0
\end{equation}

\begin{equation}
\frac{N}{2\lambda }c_{ab}-\frac{\rmi\hbar }{2}\left[ F^{-1}\right] _{ab}+\frac{1%
}{2}\Pi _{ab}=0
\end{equation}

\begin{equation}
\frac{N}{\lambda }\left\{ \bar{\chi}\left( t\right) -M^{2}\right\} -\frac{N}{%
2}\delta _{AB}G^{A1,B1}(t,t)=0
\end{equation}
where $D_{ab}\left( u,v\right) =c_{ab}\left[ \partial _{x}^{2}-\bar{\chi}%
\left( u\right) \right] \delta (u,v)$,

\begin{equation}
\Pi _{Aa,Bb}=2\frac{\delta \Gamma _{Q}}{\delta G^{Aa,Bb}};\qquad \Pi _{ab}=2%
\frac{\delta \Gamma _{Q}}{\delta F^{ab}}
\end{equation}
We shall seek a solution with the structure

\begin{equation}
G^{Aa,Bb}=\frac{\hbar }{N}\delta ^{AB}G^{ab}(u,v)  \label{eq20}
\end{equation}
which is consistent with vanishing Noether charges. Then it is convenient to
write

\begin{equation}
F^{ab}=\frac{\hbar }{N}H^{ab};\qquad \Pi _{Aa,Bb}=\delta _{AB}P_{ab};\qquad
\Pi _{ab}\left( x,y\right) =NQ_{ab}\left( x,y\right)
\end{equation}
The equations become

\begin{equation}
D_{ab}-\rmi\left[ G^{-1}\right] _{ab}+\frac{1}{N}P_{ab}=0  \label{schdy}
\end{equation}

\begin{equation}
\frac{1}{\lambda }c_{ab}-\rmi\left[ H^{-1}\right] _{ab}+Q_{ab}=0  \label{eq23}
\end{equation}

\begin{equation}
\frac{1}{\lambda }\left\{ \bar{\chi}\left( t\right) -M^{2}\right\} -\frac{%
\hbar }{2}G^{11}(t,t)=0  \label{eq24}
\end{equation}
Observe that

\begin{equation}
P_{ab}=\frac{2}{\hbar }\frac{\delta \Gamma _{Q}}{\delta G^{ab}};\qquad
Q_{ab}=\frac{2}{\hbar }\frac{\delta \Gamma _{Q}}{\delta H^{ab}}
\end{equation}
These are the exact equations we must solve. The successive $1/N$
approximations amount to different constitutive relations expressing $P_{ab}$
and $Q_{ab}$ in terms of the propagators.

For power counting in orders of $N$, observe that in any given Feynman graph each vertex
contributes a power of $N,$ each internal line a power of
$N^{-1}$, and each trace over group indices another power of $N.$
We have both $G$ and $H$ internal lines, but the $G$ lines only
appear in closed loops. On each loop, the number of vertices
equals the number of $G$ lines, so there only remains one power
of $N$ from the single trace over group labels. Therefore the
overall power of the graph is the number of $G$ loops minus the
number of $H$ lines. Now, since we only consider $2PI$ graphs,
there is a minimum number
of $H$ lines for a given number of $G$ loops. For example, if there are two $%
G$ loops, they must be connected by no less than $3$ $H$ lines,
and so this graph cannot be higher than $NNLO.$ A graph with $3$
$G$ loops cannot have less than $5$ $H$ lines, and so on.

We conclude that $\Gamma _{Q}$ vanishes at $LO,$ and therefore $%
P_{ab}=Q_{ab}=0.$ There is only one $NLO$ graph, consisting of a single $G$
loop and a single $H$ line. This graph leads to

\begin{equation}
\Gamma _{Q}^{NLO}=\left( -\rmi\hbar \right) \left( -\frac{1}{2}\right) \left( -%
\frac{N}{2Z_{0}\hbar }\right) ^{2}2N\left( \frac{\hbar }{N}\right)
^{3}c_{abc}c_{def}\int dudv\;H^{ad}\left( u,v\right) G^{be}\left( u,v\right)
G^{cf}\left( u,v\right)
\end{equation}
Therefore, we get

\begin{equation}
P_{ab}=\rmi\hbar c_{acd}c_{bef}H^{ce}G^{df}  \label{eq75}
\end{equation}

\begin{equation}
Q_{ab}=\frac{\rmi\hbar }{2}c_{acd}c_{bef}G^{ce}G^{df}  \label{eq76}
\end{equation}

%\subsection{From correlations to the reduced density matrix}

The $2PI$ $EA$ yields equations of motion for the (arbitrary time) two-point
functions of the theory. Given a solution of these equations, in principle
we may find the expectation values of a large family of composite operators
at any given time. Suppose we adopt a coarse-grained description where we
choose a certain number of these expectation values as the relevant
variables to describe the system. Then there will be a single density matrix
which has maximum von Neumann entropy with respect to the class of states
reproducing the preferred expectation values. This maximum entropy density
matrix is the reduced density matrix for the system, and its entropy its
correlation entropy. The $H$ theorem is the statement that the correlation
entropy grows in time, when the correlations themselves are evolved using
the equations derived from the $2PI$ $EA$ truncated to some order in the $1/N$
expansion. For details in the proof refer to \cite{CH03}.

%\subsection{On the Relation of nPI, Large $N$ and Loop Expansion}
%\label{sec:nPI_largeN}
%
%With the somewhat coarse impression that loop expansion provides quantum corrections and large $N$ corresponds to macro,
%one can see how nPI bear on the issues of macro and quantum by examining the relation between these three expansions. This was first discussed in \cite{CH95,CH00} in terms of the so-called master effective action.
%
%@@ recap what's there. Extract what is relevant in nPI to m/M issues and entropy/ H-theorem. The following is from Yigit's writeup:@
%
%@@ YS
%
%The relation between $nPI$, loop expansion and truncation of the Schwinger-Dyson hierarchy has been explained in section \ref{sec:entropy_slaving}. In this section we want to discuss how large N expansion fits into this picture. We begin with the observation that $N$ plays a dual role in the QMON model. On the one hand $N$ is the number of degrees of freedom. On the other hand $N$ appears in the denominator of the interaction term in the system Hamiltonian \ref{eq:Hamiltonian}. The large N limit is usually associated with the ``\emph{thermodynamic limit}" only, however for the QMON model it is the weak coupling limit as well. If we would like to make an analogy to the classical BBGKY hierarchy of section \ref{sec:BBGKY}, large N approximation would correspond to the weak coupling approximation.

\subsection{H theorem at $NLOLN$ and its implications}

It is clear that if we could solve the exact evolution, the $H$ theorem
would be manifest: Given $\rho \left( t_{0}\right) $ at the initial time $%
t_0 $, solve the exact Liouville equation up to a time $t_{1}$. Let $\bar{%
\rho}\left( t_{1}\right) $ be the result. We extract the new expectation
values from $\bar{\rho}\left( t_{1}\right) $ and use them to construct the
new maximum entropy density matrix $\rho \left( t_{1}\right) $. Then $%
S\left[ \rho \left( t_{1}\right) \right] \geq S\left[
\bar{\rho}\left( t_{1}\right) \right] ,$ by definition, and
$S\left[ \bar{\rho}\left( t_{1}\right) \right] =S\left[ \rho
\left( t_{0}\right) \right] $, because the exact evolution is
unitary, thus the $H$ theorem.

What we need to determine is whether, given the approximate
dynamics for the expectation values provided by the $1/N$ scheme,
the $H$ theorem still holds. From our first paper (MQP1) in this series \cite{MQP1} we know that to $LO$ it does
not  because $LO$ large $N$ yields a mean field theory whose equation of motion, the Vlasov equation, is unitary.
%From the angle of addressing the thermalization issue, the $LO$ approximation totally misses the point.
Thus there is no net entropy production and no $H$-theorem. In such
a case, if one works with a Fock representation, for boson
fields, the number of particles increases with time and can be
used as a measure of field entropy, what Hu and Pavon \cite{HuPav}
called an intrinsic measure of field entropy. (See also \cite{HKMP96}.) However, this should not be
confused with the correlation entropy of the Boltzmann kind under study for which the $H$%
-theorem is defined.

Of course, we expect better approximations to be closer to the
exact dynamics, therefore requiring less external control of (or
rather, tampering with) the system. This reduces the entropy loss
to the environment.
%Eventually correlation entropy production becomes the dominant factor, and a $H$ theorem is obtained.
Physically, whenever some higher order correlations are ignored, we expect to obtain an $H$ theorem because such a
description of the system is incomplete. This is the physical origin of
\textit{correlation entropy}, as explained in the Introduction.
%The system as described by a finite order $1/N$ approximation is an open system, in the following sense: Suppose we try to reproduce it in the lab.
In order to have the different correlations evolving according to the $2PI$ $1/N$ equations
appropriate to the desired order, instead of the accurate description from the full Schwinger
- Dyson hierarchy, we must keep nudging it to conform to this
artificially created condition due to our inability to
comprehend the complete picture.
%So there is energy and (in principle) entropy flow in and out of the system besides correlation entropy production. The sum of the entropy produced in the system, and that exchanged with the environment (which keeps the system on course) need not be positive.
This is the source of a new kind of noise arising from ignoring the higher order correlation functions, called correlation noise, which, when incorporated in the Schwinger-Dyson hierarchy, yields the (nth-correlation order) stochastic SD equations, as explained in \cite{CH00}.

%To summarize, the relevant question is not whether there is a $H$ theorem for a given choice of relevant variables, which is obvious, but rather if the $NLO$ approximation is good enough to make it manifest, or we must go even higher.

\subsection{$2PI$ $EA$, H-Theorem, quantum-classical and micro-Macro}
\label{sec:m-M}

We can now ask the question: How does quantum correlations get registered in the $2PI$ $EA$ and how does the existence of H-theorem at the $NLO$ expansion pertain to the quantum-Macro issue?

First, how do the quantum features show up in the $2PI$ effective action? Only equations of evolution for the first two moments can be obtained from the $2PI$ $EA$. It is known that \cite{Habib08} the equations of motion for the first two moments are identical for classical and quantum systems. %However this does not mean that there is no quantum correction in the dynamics.
Quantum corrections arise from the way the higher order moments are treated implicitly in the effective action approach. Because of this implicit dependence the $2PI$ $EA$ approach is not easy to work with in addressing the differences of quantum and classical dynamics. One could consider going to higher orders, at least 3PI, and see if the third moment equations possess the terms absent in classical dynamics at a given order in large $N$.

The micro-Macro (m-M) demarkation issue is as difficult to tackle, if not more, as the quantum-classical transition issue. Going from micro to macro requires coarse graining. QMON model is a ``democratic" system \footnote{Democratic'  is a figurative description of a system of identical particles and `autocratic'refers to a system containing a distinct member from the rest. This corresponds to the Boltzmann versus Langevin paradigms describing an effectively open (by correlation order, which depends on the level of precision of observation) versus an open (by ab initio designation) system. See, \cite{CH08}.}, hence the coarse graining should be ``democratic" as well.    The $2PI$ approach provides us with a reduced description in terms of the two lowest order Green's functions, which we shall refer to as the relevant variables. Higher order Green's functions are referred to as irrelevant variables. If one keeps track only of the  relevant variables information about irrelevant variables is lost, by choice. If these variables are coupled and one wishes to see how the relevant variables (our system) is affected by the irrelevant variables (its environment) one needs to take into account the back-action of the irrelevant variables. This back-action included reduced system (in terms of e.g., a reduced density matrix) would show dissipative dynamics and a H-theorem will obtain for the correlation entropy.

The question is whether the H-theorem holds for the  dynamics obtained at a certain order of the $1/N$ expansion. There is no H-theorem at the leading order. This can be understood by realizing that the $LO$ is equivalent to Gaussian approximation, and in this approximation correlations of order higher than two have no dynamics. As a result the mean field and the two point function contain all the available partial information because of the approximation imposed.  The fact that H-theorem holds at $NLO$ shows that the $NLO$ approximation is capable of capturing the back-action from the irrelevant variables which engenders dissipation in the dynamics of the relevant variables. Very coarsely, in the large N perspective the NLO may be viewed as providing us a witness to the micro-macro transition. Note this does not mean that $NLO$ large $N$ is a good approximation in general. Actually  we know from Mihaila et. al. \cite{MACDH00} that it fails at long times although the H-theorem holds at long times, albeit only shown for a vacuum initial state \cite{CH03}. The existence of an H-theorem, hence irreversibility, is a way of arguing that we have gone beyond the  micro-domain.

In the QMON model $N$ plays a dual role. On the one hand $N$ is the number of degrees of freedom. On the other hand $N$ appears in the denominator of the interaction term in the system Hamiltonian \ref{eq:Hamiltonian}. The large N limit is usually associated with the number of components $N\rightarrow\infty$ only, however for the QMON model it is the weak coupling limit as well. If we would like to make an analogy to the classical BBGKY hierarchy of section \ref{sec:BBGKY}, large N approximation would have the same effect as the weak coupling approximation. The fact that there is no H-theorem at $LO$ but there is an H-theorem at $NLO$ is analogous to the fact that Vlasov equation doesn't lead to entropy production, whereas the Landau equation does so. %, due to the absence/presence of the collision term.

Entropy increase is a signifier of irreversibility and irreversibility is usually associated with macroscopia. $LO$ approximation gives reversible dynamics, while at $NLO$ irreversibility emerges. This fact appears to be in contradiction with the association of irreversibility with macroscopia, since $LO$ theory is successful in describing systems with more components $N$ than $NLO$ theory. The apparent paradox can be resolved by appreciating the dual role of $N$ in the QMON model. The above example shows the interplay between system-size and interaction strength in the emergence of macroscopic qualities in a system. The system-size alone is not enough to qualify the system as macroscopic.

The work of Mihaila et. al. \cite{MACDH00} reveals a critical value of $N$ below which the $NLO$ approximation does poorly ($N_c=18.6$ for the parameters they chose for the simulations). As was discussed in MQP1 this is a way of quantifying when $N$ is actually large enough for a certain order to yield good results. This is a stronger condition than the validity of H-theorem (which holds at $NLO$ for any $N$), since it requires the dynamics to follow the exact one, rather than just one condition of nondecreasing entropy. We may combine these two results into the following interpretation: The H-theorem is necessary for a satisfactory reduced description of a macroscopic system. It holds at $NLO$. $NLO$ is a reasonable approximation for the system dynamics for $N>N_c$. Hence for $N>N_c$ the reduced description is consistent.

%%%%%%%%%%%%%%%%%%%%%%%%%%%%%%%%%%%%%%%%%%%%%%%%%%%%%%%%%%%%%%%%%%%%%%%%%%%%%%%%%%%%%%%%%%%%%%%%%%%%%%%%%%%%%%%%%%%

\section{Model study: $2PI$, $NLOLN$, loop expansions in Nonequilibrium BEC dynamics}

%Invoke results in Rey et al PRA (04) on NEq dynamics of BEC atoms to illustrate how large $N$, $2PI$ and second order enter. Note the role of correlation. (J - U effects in Bose Hubbard model)
%Strongly correlated systems:  Look into the mesoscopic literature and see if CMP people have used $2PI$. Mid-80s people working on this field (Patrick Lee, Michael Stone, Boris Alshuler et al)tackle transport with Keldysh method.

We begin with a description of the well-known Bose-Hubbard model and state our goal in this illustrated example.
We introduce the $2PI$ generating functional to construct the $2PI$ effective action ($EA$)
$\Gamma[\phi,G]$ and Green's functions. We perform a perturbative expansions on $\Gamma_2$ and define the various
approximation schemes. We then  derive the equation of motion from the $CTP$ $2PI$ $EA$  and discuss the results under each approximation scheme, starting with the Hartree-Fock-Bogoliubov (HFB) approximation followed by second order expansions which includes the 1/$\mathcal{N}$ expansion and then the full second order expansion. The results from these approximations are compared with the exact numerical answers so as to exhibit the effects of quantum correlations, the coupling strength and how the results depend on increasing $N$.

The dynamics of an ultracold \ bosonic gas in an optical lattice can be
described by a Bose-Hubbard model where the system parameters are controlled
by laser light. For a one dimensional lattice the Hamiltonian is:

\begin{equation}
\hat{H}=-J\sum_{i}(\hat{\Phi}_{i\;}^{\dagger }\hat{\Phi}_{i+1\;}+\hat{%
\Phi}_{i+1\;}^{\dagger }\hat{\Phi}_{i\;})+\sum_{i}\epsilon _{i}\hat{%
\Phi}_{i\;}^{\dagger }\hat{\Phi}_{i\;}+\frac{1}{2}U\sum_{i}\hat{\Phi%
}_{i\;}^{\dagger }\hat{\Phi}_{i\;}^{\dagger }\hat{\Phi}_{i\;}\hat{%
\Phi}_{i\;},  \label{BHH}
\end{equation}
where $\hat{\Phi}_{i}$ and $\hat{\Phi}_{i}^{\dagger }$ \ are the
annihilation and creation operators at the site $i$ \ which obey
the canonical commutation relations for bosons. Here, the
parameter $U$ denotes the strength of the on-site repulsion of two
atoms on the site $i$; the parameter $\varepsilon _{i}$ denotes
the energy offset of each lattice site due to an additional slow
varying external potential such as a magnetic trap and $J$ denotes
the hopping rate between adjacent sites. Because the
next-to-nearest neighbor amplitudes are \ typically two orders of
magnitude smaller, tunneling to them \ can be neglected. We denote the total number of atoms by $N$ and the number of lattice sites by I. Only a one-dimensional homogeneous lattice with
periodic boundary conditions is considered.

A dimensionless parameter that is convenient to describe
the different regimes of H  is the coupling strength $\lambda
\equiv NU/ IJ$. Different from a homogeneous system without a
lattice where at zero temperature the superfluid fraction is
always unity, the presence of the lattice changes the superfluid
properties and even at zero temperature, the superfluid
fraction decreases with the lattice depth. For strong-coupling
strengths \cite{Fishers} $\l > \l_{crit}$,
\be
\l_{crit} \sim \frac{2N}{I^2} [2N + I + \sqrt{(2N + I)^2 + I^2}],
\ee
it is known that the ground state undergoes a quantum phase
transition from a superfluid to a Mott insulator.
In the weakly interacting regime, $\l \ll 1$, where tunneling
overwhelms the repulsion, to a good approximation quantum
fluctuations can be neglected and the properties of the system
can be described by replacing the operator on the lattice site
i by a classical c number. It can be said that most of the
atoms are in the zero quasimomentum state.
In the intermediate regime $1 \ll \l \ll \l_{crit} /2$ the interactions
between the bosons can be very strong but the ground state is
nevertheless a superfluid. For these interaction parameters a
self-consistent HFB-Popov theory gives a good description
of the system. However, different from the weak interacting
regime where the depletion of the zero quasimomentum state
is very small and has  little effect on the superfluid properties,
in this intermediate regime, depleted atoms spread over
the central part of the band and reduces the superfluid fraction.
As interactions are further increased the depleted population
completely fills the band and cancels the superfluid
properties. The system reaches the Mott insulator regime, where atoms are  localized at each lattice site and the eigenstates of the system are almost Fock states with vanishing number fluctuations per lattice site. The dynamics in the intermediate regime is the focus of study in \cite{Rey04}, where the superfluid properties are important but quantum fluctuations cannot be ignored.

\subsection{$2PI$ $EA$ $\Gamma [\protect\phi ,G]$}

The first requirement for the study of nonequilibrium processes is a general
initial-value formulation depicting the dynamics of interacting quantum
fields. The $CTP$ or Schwinger-Keldysh effective action formalism \cite{ctp}
serves this purpose. The second requirement is to describe the evolution of
the correlation functions and the mean field on an equal footing. The two
particle \ irreducible ($2PI$) formalism \cite{2pi} where the correlation \
functions appear also as independent variables\ serves this purpose. By
requiring the generalized (master) $CTP$ effective action \cite{mea} to be
stationary with respect to variations \ of the correlation functions \ an
infinite set of coupled (Schwinger-Dyson) equations for the correlation
functions is obtained which is a quantum analog of the BBGKY hierarchy. The
$2PI$ effective action produces two such functions in this hierarchy.
%In this section we shall focus on the $2PI$ formalism, and then upgrade it to the $CTP$ version in the next section.

The classical action corresponding to the Bose-Hubbard Hamiltonian
(\ref{BHH}) is given in terms of the
complex fields $\Phi _{i}^{{}}$ and $\Phi _{i}^{\ast }$ by

\begin{equation}
S[\Phi _{i}^{\ast },\Phi _{i}]=\int dt\sum_{i}\left( \rmi \hbar \Phi
_{i}^{\ast }(t)\partial _{t}\Phi _{i}(t)+J\left( \Phi _{i}^{\ast
}(t)\Phi _{i+1}(t)+\Phi _{i}(t)\Phi _{i+1}^{\ast }(t)\right) -\frac{%
U}{2}\Phi _{i}^{\ast }(t)\Phi _{i}^{\ast }(t)\Phi _{i}(t)\Phi
_{i}(t)\right),
\end{equation}
where, as before, $i$ denotes the lattice position, $\ J$ is the hopping
rate and $U$ is the interaction strength. We limit the analysis to the case
when no external potential is present and include only nearest neighbor
hopping. To compactify our notation we introduce $\Phi _{i}^{a} (a=1,2)$
defined by
\begin{equation}
\Phi _{i} =\Phi _{i}^{1}, \;\; \Phi _{i}^{\ast } =\Phi _{i}^{2}.
\end{equation}
In terms of these fields the classical action takes the form

\begin{equation}
\label{classAction}
S[\Phi ]=\int dt\sum_{i}\left( \frac{1}{2}h_{ab}\Phi _{i}^{a}(t)\hbar
\partial _{t}\Phi _{i}^{b}(t)+J\sigma _{ab}\Phi _{i+1}^{a}(t)\Phi
_{i}^{b}(t)-\frac{U}{4\mathcal{N}}(\sigma _{ab}\Phi
_{i}^{a}(t)\Phi _{i}^{b}(t))^{2}\right) ,
\end{equation}
where $\mathcal{N}$ is the number of fields (two in this case), and summation over repeated field indices $%
a,b=(1,2)$ is implied. $h_{ab}$ and $\sigma _{ab}$ are matrices defined as

\begin{equation}
h_{ab}=\rmi\left(
\begin{array}{cc}
0 & -1 \\
1 & 0
\end{array}
\right) \qquad \sigma _{ab}=\left(
\begin{array}{cc}
0 & 1 \\
1 & 0
\end{array}
\right)
\end{equation}

After second quantization the fields $\Phi _{i}^{a}$ are promoted
to operators. We denote the expectation value of the field
operator or mean field \ by $\phi _{i}^{a}(t)$ and the expectation
value\ of the composite field \ by $G_{ij}^{ab}(t,t^{\prime })$.
Physically, $|\phi _{i}^{a}(t)|^{2}$ is the condensate population
and the composite fields determine the fluctuactions around the
mean field.

\begin{eqnarray}
\phi _{i}^{a}(t) &=&\left\langle \Phi _{i}^{a}(t)\right\rangle, \\
\hbar G_{ij}^{ab}(t,t^{\prime }) &=&\left\langle T_{C}\Phi
_{i}^{a}(t)\Phi _{i}^{b}(t^{\prime })\right\rangle -\left\langle
\Phi _{i}^{a}(t)\right\rangle \left\langle \Phi _{i}^{b}(t^{\prime
})\right\rangle.
\end{eqnarray}
The brackets denote taking the expectation value with respect to the density matrix and $%
T_{C}$ denotes time ordering along a contour C in the complex plane.

\bigskip

All correlation functions of the quantum theory can be obtained from the
effective action $\Gamma \lbrack \phi ,G]$, the two particle irreducible
generating functional \ for Green's functions parametrized by $\phi
_{i}^{a}(t)$ and the composite field $G_{ij}^{ab}(t,t^{\prime })$. To get an
expression for the effective \ action we first define \ the functional $Z[%
\mathbf{J,K}]$ \cite{2pi} as

\begin{eqnarray}
Z[\mathbf{J,K}] &=&e^{\rmi/\hbar W[\mathbf{J,K}]}  \label{zeq} \\
&=&\prod_{a}\int D\Phi ^{a}\exp \left\{ \frac{\rmi}{\hbar }\left(
S[\Phi ]+\int dt\sum_{i}\mathbf{J}_{ia}(t)\Phi
_{i}^{a}(t)+\frac{1}{2}\int
dtdt^{\prime }\sum_{ij}\Phi _{i}^{a}(t)\Phi _{j}^{b}(t^{\prime })%
\mathbf{K}_{ijab}(t,t^{\prime })\right) \right\} ,  \nn
\end{eqnarray}
where we have introduced the following index lowering convention
\begin{equation}
X_{a}=\sigma _{ab}X^{b}.
\end{equation}
The functional integral (\ref{zeq}) is a sum over classical histories of the
field $\Phi _{i}^{a}$ \ in the presence of \ the local source \ $\mathbf{J%
}_{ia}$ and the non local source $\mathbf{K}_{ijab}.$ The coherent
state measure is included in $D\Phi $. The addition of \ the
two-particle source term is what characterizes the $2PI$ formalism.

We define $\Gamma \lbrack \phi ,G]$ as the double Legendre transform of $W[%
\mathbf{J,K}]$ such that

\begin{eqnarray}
\frac{\delta W[\mathbf{J,K}]}{\delta \mathbf{J}_{ia}(t)} &=&\phi _{i}^{a}(t),
\\
\frac{\delta W[\mathbf{J,K}]}{\delta \mathbf{K}_{ij ab}(t,t^{\prime })} &=&%
\frac{1}{2}[\phi _{i}^{a}(t)\phi _{i}^{b}(t^{\prime })+\hbar
G_{ij}^{ab}(t,t^{\prime })].
\end{eqnarray}
Expressing $\mathbf{J}$ and $\mathbf{K}$ in terms of $\phi $ and $G$ \ yields

\begin{eqnarray}
\label{eq:yigit}
\Gamma \lbrack \phi ,G] &=&W[\textbf{J},\textbf{K}]-\int
dt\sum_{i}\mathbf{J}_{ia}(t)\phi _{i}^{a}(t)-\frac{1}{2}\int
dtdt^{\prime }\sum_{ij}\phi _{i}^{a}(t)\phi
_{j}^{b}(t^{\prime })\mathbf{K}_{ijab}(t,t^{\prime }) \\
&&-\frac{\hbar }{2}\int dtdt^{\prime }\sum_{ij}G_{ij}^{ab}(t,t^{\prime })%
\mathbf{K}_{ijab}(t,t^{\prime }). \label{gamm} \nn
\end{eqnarray}
{}From this equation the following identity can be derived:

\begin{eqnarray}
\frac{\delta \Gamma \lbrack \phi ,G]}{\delta \phi _{i}^{a}(t)} &=&-\left(
\mathbf{J}_{ia}(t)+\int dt^{\prime }\sum_{j}(\mathbf{K}_{ijad}(t,t^{\prime
}))\phi _{j}^{d}(t^{\prime })\right) ,  \label{eaeq1} \\
\frac{\delta \Gamma \lbrack \phi ,G]}{\delta G_{ij}^{ab}(t,t^{\prime })} &=&-%
\frac{\hbar }{2}\mathbf{K}_{ijab}(t,t^{\prime }).  \label{eaeq2}
\end{eqnarray}

In order to get an expression for $\Gamma \lbrack \phi ,G]$ notice
that by using (\ref{zeq}) for $W[\textbf{J}, \textbf{K}]$ and
placing it in (\ref{eq:yigit}) for $\Gamma \lbrack \phi ,G]$, it
can be written as

\begin{eqnarray}
\exp \left( \frac{\rmi}{\hbar }\Gamma \lbrack \phi ,G]\right)  &=&\prod_{a}\int
D\Phi ^{a}\exp \left\{ \frac{\rmi}{\hbar }\left( S[\Phi ]+\int dt_{i}\;%
\mathbf{J}_{ia}(t)\left[ \Phi _{i}^{a}(t)-\phi _{i}^{a}(t)\right]
\right.
\right.  \\
&&\left. \left. +\frac{1}{2}\int dt_{i}dt_{j}^{\prime }\left( \Phi
_{i}^{a}(t)\mathbf{K}_{ijab}^{{}}(t,t^{\prime })\Phi
_{j}^{b}(t^{\prime })-\phi
_{i}^{a}(t)\mathbf{K}_{ijab}^{{}}(t,t^{\prime })\phi
_{j}^{b}(t^{\prime })\right) -\frac{\hbar }{2}\Tr G\mathbf{K}\right)
\right\}
\nn \\
&=&\prod_{a}\int D\Phi ^{a}\exp \left\{ \frac{\rmi}{\hbar }\left(
S[\Phi
]-\int dt_{i}\;\frac{\delta \Gamma \lbrack \phi ,G]}{\delta \phi _{i}^{a}(t)}%
\left[ \Phi _{i}^{a}(t)-\phi _{i}^{a}(t)\right] \right. \right.
\nn
\\
&&\left. \left. -\frac{1}{\hbar }\int dt_{i}dt_{j}^{\prime
}\;\left[ \Phi
_{i}^{a}(t)-\phi _{i}^{a}(t)\right] \frac{\delta \Gamma \lbrack \phi ,G]}{%
\delta G_{ij}^{ab}(t,t^{\prime })}\left[ \Phi _{i}^{b}(t^{\prime
})-\phi
_{i}^{b}(t^{\prime })\right] +\Tr G\frac{\delta \Gamma \lbrack \phi ,G]}{%
\delta G}\right) \right\} ,  \nn
\end{eqnarray}
where Tr means taking the trace. For simplicity we have denoted
$\int dt\sum_{i}$ by $\int dt_{i}$. Defining the fluctuation
field, $\varphi_{i}^{a}=\Phi_{i}^{a}-\phi_{i}^{a}$, we have

\begin{eqnarray}
\Gamma \lbrack \phi ,G]&-&\Tr G\frac{\delta \Gamma \lbrack \phi ,G]}{\delta G}
=-\rmi\hbar \ln \prod_{a}\int D\varphi ^{a}\exp \left( \frac{\rmi}{\hbar }S[\phi
,G;\varphi ]\right)  \label{eact} \\
\fl S[\phi ,G;\varphi ] &=&S[ \phi +\varphi]-\int dt_{i}\;\frac{\delta
\Gamma \lbrack \phi ,G]}{\delta \phi _{i}^{a}(t)}\varphi
_{i}^{a}(t)-\frac{1}{\hbar }\int dt_{i}dt_{j}^{\prime }\;\varphi
_{i}^{a}(t)\frac{\delta \Gamma \lbrack \phi ,G]}{\delta
G_{ij}^{ab}(t,t^{\prime })}\varphi _{i}^{b}(t^{\prime }).
\label{shift}
\end{eqnarray}
By introducing the classical inverse propagator $iD^{-1}(\phi )$ given by

\begin{eqnarray}
\rmi D_{ijab}(t,t^{\prime })\;^{-1} &=&\frac{\delta S[\phi ]}{\delta \phi
_{i}^{a}(t)\delta \phi _{j}^{b}(t^{\prime })} \\
&=&\left( \delta _{ij}h_{ab}\partial _{t}+J(\delta _{i+1j}+\delta
_{i-1j})\sigma _{ab}\right) \delta (t-t^{\prime })  \nn \\
&&-\frac{U}{\mathcal{N}}\left( 2\phi _{ia}(t)\phi _{ib}(t)+\sigma _{ab}\phi
_{i}^{c}(t)\phi _{ic}(t)\right) \delta _{ij}\delta (t-t^{\prime }),  \nn
\end{eqnarray}
the solution of the functional integro-differential equation (\ref{eact})
can be expressed as

\begin{equation}
\Gamma \lbrack \phi ,G]=S[\phi ]+\frac{\rmi}{2}\Tr\ln G^{-1}+\frac{\rmi}{2}%
\Tr D^{-1}(\phi )G+\Gamma _{2}[\phi ,G]+const.  \label{2pi}
\end{equation}
The quantity $\Gamma _{2}[\phi ,G]$ is conveniently \ described in
terms \ of the diagrams generated by the \ interaction terms in
$S[ \phi +\varphi]$ which are \ of cubic and higher orders in
$\varphi $.

\begin{equation}
S_{int}[ \phi +\varphi]=-\frac{U}{4\mathcal{N}}\int
dt_{i}\;(\varphi _{ib}(t)\varphi
_{i}^{b}(t))^{2}-\frac{U}{\mathcal{N}}\int dt_{i}\;\varphi
_{i}^{a}(t)\phi _{ia}(t)\varphi _{i}^{b}(t)\varphi _{ib}(t).
\end{equation}
It consists of all \ two-particle irreducible vacuum graphs \ (the
diagrams representing these interactions do not become
disconnected by cutting two propagator lines) in the theory with
propagators set equal to $G$ and vertices determined by \ the \
interaction terms in $S[ \phi +\varphi]$ .

Since physical processes correspond to \ vanishing sources $\mathbf{J}$ and $%
\mathbf{K}$, \ the dynamical equations of motion for the mean field and the
propagators are found \ by using the expression (\ref{2pi}) in equations (%
\ref{eaeq1})\ and (\ref{eaeq2}), and setting the right hand side equal to
zero. This procedure leads to the following equations:

\begin{eqnarray}
h_{ab}\hbar \partial _{t}\phi _{i}^{b}(t) &=&-J(\phi _{i+1a}(t)+\phi
_{i-1a}(t))+\frac{U}{\mathcal{N}}(\phi _{id}(t)\phi
_{i}^{d}(t)+G_{ii\;c}^{\;c}(t,t))\phi _{ia}(t)+  \label{cond} \\
&&\frac{U}{\mathcal{N}}((G_{iiad}(t,t)+G_{iida}(t,t))\phi _{i}^{d}(t)-\frac{%
\delta \Gamma _{2}[\phi ,G]}{\delta \phi _{i}^{a}(t)},  \nn
\end{eqnarray}

\noindent and

\begin{eqnarray}
G_{ijab}^{-1}(t,t^{\prime }) &=&D_{ijab}(t,t^{\prime })^{-1}-\Sigma
_{ijab}(t,t^{\prime }),  \label{inv} \\
\Sigma _{ijab}(t,t^{\prime }) &\equiv &2 \rmi \frac{\delta \Gamma _{2}[\phi ,G]}{%
\delta G_{ij}^{ab}(t,t^{\prime })} .  \label{pder}
\end{eqnarray}
Equation\ (\ref{inv}) can \ be rewritten \ as a partial differential
equation \ suitable for initial value problems \ by convolution with $G$.
This differential equation reads explicitly

\begin{eqnarray}
h_{c}^{a}\hbar \partial _{t}G_{ij\;}^{\;cb}(t,t^{\prime })
&=&-J(G_{i+1j}^{ab}(t,t^{\prime })+G_{i-1j}^{ab}(t,t^{\prime }))+\frac{U}{%
\mathcal{N}}(\phi _{id}(t)\phi _{i}^{d}(t))G_{ij}^{ab}(t,t^{\prime })+
\label{fluc} \\
&&\frac{2U}{\mathcal{N}}\phi _{i}^{a}(t)G_{ij}^{cb}(t,t^{\prime })\phi
_{ic}(t)+\rmi\int dt_{k}^{\prime \prime }\Sigma _{ikc}^{\;a}(t,t^{\prime \prime
})G_{kj}^{cb}(t^{\prime \prime },t^{\prime })+\rmi\delta ^{ab}\delta
_{ij}\delta (t-t^{\prime }).  \nn
\end{eqnarray}
The evolution of $\phi ^{a}$ and $G^{ab}$ is determined by Eqs. (\ref{cond})
and (\ref{fluc}) once $\Gamma _{2}[\phi ,G]$ is specified.

\subsection{Perturbative Expansion of $\Gamma _{2}(\protect\phi ,G)$ and
Approximation Schemes}

\bigskip The diagrammatic expansion of \ $\Gamma _{2}$ is illustrated in the
accompanying Figure where two and three-loop vacuum diagrams are shown.

\begin{figure}[tbh]
\begin{center}
\includegraphics[width=4.2 in]{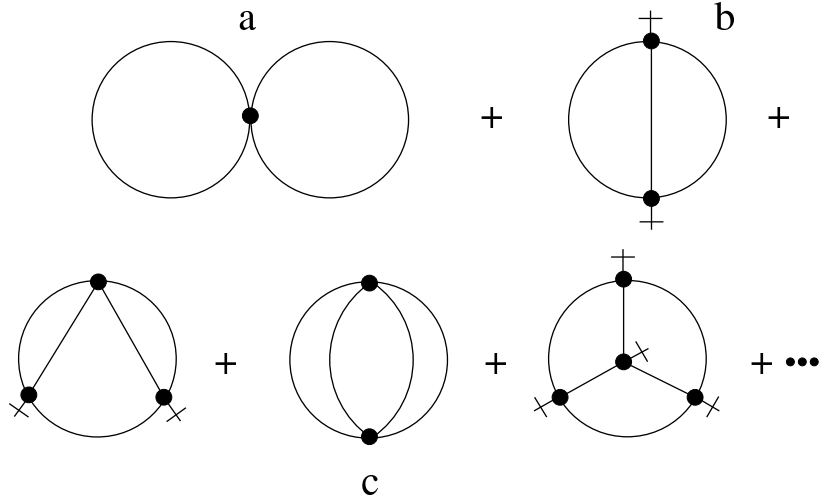}
\end{center}
\caption{Two-loop (upper row) and three-loop diagrams (lower row)
contributing to the effective action. Explicitly, the diagram a is what we
call the \emph{double-bubble} , b the \emph{setting-sun} and c the \emph{%
basket-ball}. }
\label{Fig2}
\end{figure}

The dots where four lines meet represent interaction vertices. \ The expression \
corresponding to each vacuum diagram should be multiplied by a factor $%
(-i)^{l}(i)^{s-2}$ where $l$ is the number of solid lines and \ s the number
of loops the diagram contains.

The action \ $\Gamma $ including the full diagrammatic series for $\Gamma
_{2}$ gives the full dynamics. It is of course not feasible to obtain an
exact expression for $\Gamma _{2}$ in a closed form. Various approximations
for the full $2PI$ effective action can be obtained by truncating the
diagrammatic expansion \ for $\Gamma _{2}$. Which approximation is most
appropriate depends on the physical problem under consideration:

\subsection{The standard approaches}
\label{sec:standard_approaches}

\begin{enumerate}
\item  \ Bogoliubov (One-loop) Approximation:

The simplest approximation consists of discarding $\Gamma _{2}$ altogether.
This yields the so called Bogoliubov or \ one-loop approximation whose
limitations have been extensively documented in the literature (\cite
{Morgan,BurnettComparisons}).

\item  Time-dependent Hartree-Fock-Bogoliubov (HFB) Approximation:

A truncation of $\Gamma _{2}$ retaining only the first order diagram in U,
i.e., keeping only the \emph{double- bubble} diagram, Fig. $a$, yields
equations of motion of $\phi $ and $G$ which correspond to the time
dependent Hartree-Fock-Bogoliubov (HFB) approximation. This approximation
violates Goldstone's theorem, but conserves energy and particle number \cite
{Martin,Griffin,Giorgini}). The HFB equations can also be obtained by using
cumulant expansions up to the second order \cite{Burnett} in which all
cumulants containing three or four field operators \ are neglected. The HFB
approximation neglects multiple scattering. It can be interpreted as an
expansion in terms of $Ut/J,$ (where $t$ is the time of evolution) and is
good for the description of \ short time dynamics or weak interaction
strengths. %It will be described in Sec. 6.
\end{enumerate}

\begin{enumerate}
\item  Second Order expansion:

A truncation retaining diagrams of second order in $U$ containing besides
the \emph{double-bubble}, also the \emph{setting-sun} and the \emph{%
basket-ball}. By including the \emph{setting-sun} and the \emph{%
Basket-ball} in the approximations we are taking into account two particle
scattering processes \cite{CHR}. Second order terms lead to integro-
differential equations which depend on the time history of the system.

\item  Large-$\mathcal{N}$ approximation

The 1/$\mathcal{N}$ expansion is a controlled \ non-perturbative
approximation \ scheme which \ can be used to study non-equilibrium quantum
field \ dynamics in the regime \ of strong \ interactions\cite{AABBS02}%
. In the large $\mathcal{N}$ approach the field is modeled by $\mathcal{N}$
fields and the quantum field generating functional is expanded \ in powers
of 1/$\mathcal{N}$. \ In this sense the method is a controlled expansion in
a small parameter but \ unlike perturbation theory in the coupling constant,
which corresponds to an expansion around the vacuum, the large $\mathcal{N}$
expansion corresponds to an expansion of the theory about a strong
quasiclassical field.\
\end{enumerate}

In \cite{Rey04}  numerical solutions to the equations of motion for a moderate number of atoms
and wells up to the second order in the coupling constant $U$ were given. These exact many body solutions are useful for  determining the range of validity of the three types of approximations described above, namely, a) the HFB, b) the full second order and c) $NLO$ large $\mathcal{N}$ expansion up to second order in $U$ (the shorthands HFB, 2nd and 1/$\mathcal{N}$ respectively are used in the figure).  For the effect of the total number of atoms on the dynamics Rey et al have given the numerical solutions for a
double well system with fix ratio $UN/J=4$ and three different values of $N$:
$N$=20, 40 and 80. The results are given for the evolution of the
atomic population per well (Fig. 5), the condensate population per well and
total condensate population (Fig. 6) and the quasi-momentum intensities (
Fig. 7). (To make the comparisons easier  the numerical results
obtained for the three different values of $N$ are scaled by dividing them by the total
number of atoms. In this way for all the cases we start with an atomic
population of magnitude one in the initial populated well.)

In the exact dynamics we see that as the number of atoms is
increased the damping effects occur at  slower rates. This feature
can be noticed in the quantum dynamics of all of the observables
depicted in the Figures 5 to 7 of \cite{Rey04}. The decrease of the damping rates as
the number of atoms is increased is not surprising because
changing the number of atoms alters the quantum coherence
properties of the system. As shown in reference \cite{Walls}, the
collapse time of the condensate population is approximately given by $%
t_{coll}\sim \frac{t_{rev}}{\sigma }$ where $\sigma $ is the variance of the
initial atomic distribution and ${t_{rev}}$ is the revival time which
depends on the detailed spectrum for the Hamiltonian. In the kinetic energy
dominated regime ${t_{rev}}\sim hN/J$ (see \cite{Smerzi}), thus $%
t_{coll}\sim \frac{N}{\sigma }$. For our initial conditions the variance is
proportional to $\sigma =\sqrt{N}$ so $t_{coll}\sim \sqrt{N}$. Besides
damping rates, the qualitative behavior of the exact quantum dynamics is not
affected very much as the number of atoms is increased for a fixed $UN/J$.

The improvement of the $2PI$ approximations as $N$ is increased, as a
result of the increase in the initial number of coherent atoms is
in fact observed in the plots. Even though the problem of
underdamping in the HFB approximation and the overdamping in the
second order approaches are not cured, as the number of atoms is
increased, we do observe a better matching with the full quantal
solution. The 1/$\mathcal{N}$ expansion shows the fastest
convergence, as can be more easily observed in the
quasi-momentum distribution plots, Fig. 7. The better agreement of the 1/$%
\mathcal{N}$ expansion relies on the fact that even though the number of
fields is only two in our calculations the 1/$\mathcal{N}$ expansion is an
expansion about a strong quasiclassical field configuration, a feature we have explained in MQP1.\\

\noindent{\bf Part II: Effective Scale and Infrared Dimension}
\vskip .5cm

\section{Effective field theory, coupling and the effective scale of a quantum system}

Effective field theory is a very powerful tool in quantum field
theory, and in particular it gives a new point of view about the
meaning of renormalization.

Consider a quantum field theory with a characteristic energy scale
$E_0$, and suppose that we are interested in physics at some lower
energy scale $E$. The basic idea of effective field theory is that
one can choose a cutoff energy scale $\Lambda$ at or slightly
below $E_0$, and divide the fields in the path integral into low-
and high- frequency parts, $\phi = \phi_L + \phi_H$, where $\phi_L
(\phi_H)$ are the field modes with frequencies lower(higher) than
$\Lambda$. Now integrate over the  high frequency fields $\phi_H$
\begin{eqnarray}
\int D\phi_H D\phi_L e^{i S(\phi_H,\phi_L)} = \int  D\phi_L e^{i
S_{\Lambda}(\phi_L)}
\end{eqnarray}
where $e^{i S_{\Lambda}(\phi_L)} = \int D\phi_H  e^{i
S(\phi_H,\phi_L)}$. The $S_{\Lambda}(\phi_L)$ is called the
low-energy or Wilsonian effective action. It's different from the
1PI effective action which are generated from integrating over all
frequencies but keeping only 1PI graphs.

One can expand the low-energy effective action in terms of local
operators which are consistent with the symmetries of the problem,
\begin{eqnarray}
 S_{\Lambda}(\phi_L) = \int  d^D x \sum_i g_i \hat{Q}_i.
\end{eqnarray}
In units of $\hbar=c=1$, if an operator $\hat{O}_i$ has unit
$E^{d_i}$ where $d_i$ is the dimension of operator $\hat{O}_i$
then the parameter $g_i$ has unit $E^{D-d_i}$ where $D$ is the
spacetime dimension. In the case of a free scalar field in $D$
dimensions, with action
\begin{eqnarray}
 S = \int  d^D x \left( \frac{1}{2}\partial_\mu \phi \partial^\mu \phi - \frac{1}{2}m^2 \phi^2\right).
\end{eqnarray}
The action $S$ is dimensionless, so the dimension of $\phi(x)$ is
determined from the kinetic term to be $[\phi]=\frac{D-2}{2}$ and
the dimension of $m^2$ is $[m^2]=2$. An operator $\hat{O}_i$
constructed from $M$ $\phi's$ and $N$ derivatives has dimension
$d_i= M(\frac{D-2}{2}) + N$. Define dimensionless couplings
$\lambda_i = \Lambda^{d_i-D} g_i$. Now for a process at energy
scale $E$, one can dimensionally estimate the contribution from the
$i$-th term of the effective action
\begin{eqnarray}
 \int  d^D x g_i \hat{O}_i \sim \lambda_i (\frac{E}{\Lambda})^{d_i -D}.
\end{eqnarray}
If $d_i > D$, this term will become less and less important at
lower and lower energies, and is called an irrelevant operator. If
$d_i < D$, the operator is more and more important at lower and
lower energies and is called a relevant operator. If $d_i = D$ the
operator is equally important at all scales and called marginal.
In most cases there is only a finite number of relevant and
marginal operators so the low energy physics depends only on a
finite number of parameters.

The low energy physics depends on the short distance theory only
through the relevant and marginal couplings, and possibly through
some leading irrelevant couplings if one measures small enough
effects.

It should be pointed out that the coupling strength measured will
vary with energy(scale). This can easily be seen from the point of
view of effective field theory. We could write the path integral
as
\begin{eqnarray}
\int_{k\leq \Lambda} D\phi_k  e^{i S_{\Lambda}(\phi_k)} =
\int_{k'\leq \Lambda^{\prime}} D\phi_{k^{\prime}} e^{i
S_{\Lambda^{\prime}}^{\prime}(\phi_{k^{\prime}})}.
\end{eqnarray}
The action $S_{\Lambda}(\phi_k)$ contains all momentum modes up to
some maximum value $\Lambda$, whereas the new effective action
$S_{\Lambda^{\prime}}^{\prime}(\phi_{k^{\prime}})$ contains
momentum modes up to $\Lambda^{\prime} < \Lambda$. If one take
$\Lambda^{\prime}$ to be infinitesimally smaller than $\Lambda$,
$S_{\Lambda^{\prime}}^{\prime}$ will be infinitesimally different
from $S_{\Lambda}$. This generates a differential equation, the
Wilson equation
\begin{eqnarray}
\frac{\partial S_{\Lambda}}{ \partial \Lambda} = F(S_{\Lambda}),
\end{eqnarray}
where $F$ is some functional of the action. The Wilson equation is
the renormalization group flow equation in an infinite dimensional
space and it gives the change in action as a function of cutoff.
Since
\begin{eqnarray}
 S_{\Lambda}(\phi) = \int  d^D x \sum_i g_i \hat{Q}_i,
\end{eqnarray}
the Wilson equation then gives
\begin{eqnarray}
\frac{\partial g_i}{ \partial \Lambda} = F(\{ g_i\}).
\end{eqnarray}
If one linearizes a solution of the equation, irrelevant operators
will correspond to those directions with negative eigenvalues
whereas the relevant operators corresponds to those directions
with positive eigenvalues.

%\newpage

\section{Infrared behavior of quantum fields: Finite-size / Topology effects}

In the above we pointed out that the infrared or large scale behavior of a system depends on many factors: a) Interaction strength: difference between a free theory (Gaussian) and an interacting theory, and for an interacting theory the Gaussian fixed point as different from non-Gaussian ones, relevant versus irrelevant fixed points; b) Coupling strength which runs with energy via the RG equation. All this show the intricacy of the notion of `large scale' which we need to bear in mind in describing `macroscopic' quantum behavior. In this section, we demonstrate how the infrared behavior of a quantum system can be altered qualitatively by the geometry and topology of space, or more precisely, the `finite size' effect. This can be seen from analysis based on rather general arguments, e.g.,  on whether there exist a band structure in the spectrum of eigenvalues of the invariant operator of the order parameter field. If so there is a dimensional reduction effect where the IR behavior of the system becomes that of a lower dimension. What this conveys is that, for a quantum system near the critical point where its correlation length increases to (near) infinity its  large scale or IR behavior or macroscopic features can become sensitive to the size and topology of the underlying spacetime. Cosmological phase transition is an extreme yet valid, even realistic, manifestation of this feature.
%look like it is in a lower dimension. here the large scale behavior fall into universality classes,
A more rigorous and complete treatment of this effect is given in Hu and O'Connor \cite{HuOC86} from which our present exposition is adapted.

Consider an $N$-component self-interacting scalar field $\Phi^a
(a=1, \cdot \cdot \cdot,N)$ on a manifold of dimension D, coupled to
the background spacetime with curvature R and coupling constant $\xi$
(conformal coupling for $\xi =0$ and minimal coupling for $\xi =1$)
described by the action
\begin{equation}
S[\Phi] = \int d^Dx\sqrt{g} \left[ \frac{1}{2} \Phi^a \nabla^2 \Phi^a
+ \frac{1}{2}M_\xi^2 \Phi^2 + \frac{\lambda}{4!} \Phi^4 \right]
\; , \label{IRON1}
\end{equation}
where $M^2_{\xi} =m^2+(1-\xi)R/6$ and $\nabla^2 = - \sqrt{g}
\partial_\mu (\sqrt{g} g^{\mu \nu}\partial_\nu)$ is the
Laplace-Beltrami operator for scalar fields.  For convenience we
develop the formalism here in Euclidean space  version with signature (+,+,+,+).  We begin
with a decomposition of the field $\Phi^a$
into a background field $\bar{\phi}^a$ and a fluctuation field
$\varphi^a$ , i.e., $\Phi^a = \bar{\phi}^a + \varphi^a$.  The
background field $\bar{\phi}^a$ is required to satisfy the
classical equations of motion with an arbitrary external source.
Such a shift eliminates the linear term in the fluctuation field,
which is equivalent to performing a Legendre transform.  The
resultant action is
$$
S[\bar{\phi},\varphi]=S[\bar{\phi}]+\int d^Dx\sqrt{g}$$
\begin{equation}
\times \left\{\frac{1}{2} \varphi^a \left[\left(\nabla^2
+M_\xi^2+\frac{\lambda}{6} \bar{\phi}^2\right)
\delta^{ab}+\frac{\lambda}{3} \bar{\phi}^a \bar{\phi}^b\right]
\varphi^b + \frac{\lambda}{6}\bar{\phi}^a\varphi^a\varphi^2 +
\frac{\lambda}{4!} \varphi^4 \right\}. \label{IRON2}
\end{equation}
In the Feynman diagrammatics there are two kinds of vertices: a four point vertex
proportional to $\lambda$ and a three point vertex proportional
to $\lambda\bar{\phi}_{a}(x)$.

The effective action $\Gamma[\bar{\phi}]$ is obtained by
functionally integrating over the fluctuation fields:
\begin{equation}
e^{-\Gamma [\bar{\phi}]} = \int
[d\varphi]e^{-S[\bar{\phi},\varphi]}. \label{IRON3}
\end{equation}
The wave operator $A^{ab}$ for the fluctuating field is given by
\begin{equation}
A^{ab}= (\nabla^2+M_1^2 )
\frac{\bar{\phi}^a\bar{\phi}^b}{\bar{\phi}^2} + ( \nabla^2+M_2^2)
\left(\delta^{ab} -
\frac{\bar{\phi}^a\bar{\phi}^b}{\bar{\phi}^2}\right) \label{IRON4}
\end{equation}
where
\begin{equation}
M_1^2=M^2_{\xi}+\frac{\lambda}{2} \bar{\phi}^2, \; \; \;
M^2_2=M^2_{\xi}+\frac{\lambda}{6} \bar{\phi}^2 \; . \label{IRON6}
\end{equation}
Here $\bar{\phi}^a\bar{\phi}^b/\bar{\phi}^2$ and
$\delta^{ab}-\bar{\phi}^a\bar{\phi}^b/\bar{\phi}^2$  are
orthogonal projectors, the former projects along the direction in
the internal space picked out by $\bar{\phi}^a$ and the latter
projects into an ($N$-1)-dimensional subspace orthogonal to the
direction of $\bar{\phi}^a$. Note that the operator $\nabla^2$ does
not commute with the projectors unless $\bar{\phi}^a$ is a
constant.

When the direction in group space picked out by $\bar{\phi}^a$
does not vary from point to point around the manifold (this does
not necessarily imply $\bar{\phi}^2$ is a constant) %ref.27
the Green's function for $A^{ab}$ is given by
\begin{equation}
G^{ab}=G_1 \frac{\bar{\phi}^a \bar{\phi}^b}{\bar{\phi}^2} +G_2
\left( \delta^{ab}-\frac{\bar{\phi}^a\bar{\phi}^b}{\bar{\phi}^2}
\right)\;, \label{IRON5}
\end{equation}
where the $G_i ~ (i=1,2)$ are the Green's functions for the
operators $\nabla^2+M^2_i$

The one-loop correction is related to the sum of the logarithms of
the determinants of the fluctuation operators.  So in this case,
by using the projection operators, the one-loop effective action
is given by $$ \Gamma[\bar{\phi}]=S[\bar{\phi}]-\frac{1}{2}\Tr\ln
G_1-\frac{N-1}{2}\Tr \ln G_2 $$
\begin{equation}
=S[\bar{\phi}]+\frac{1}{2} \sum_{l}d_l\ln \lambda_{1l} +
\frac{N-1}{2} \sum_{l} d_l\ln  \lambda_{2l} \;, \label{IRON7}
\end{equation}
where $S[\bar{\phi}]$ is the classical action, $\lambda_{il}$ and
$d_l$ are the eigenvalues and degeneracies of the operators $\nabla^2 + M^2_i$.  % to use the example of the one-component theory.  The effective action is then given by (\ref{IRON7}) without the last term.

The determinant of A is formally divergent and needs to be
regularized.  There are a number of commonly used regularization
methods. %ref.1
If the space is Riemannian and has sufficient symmetry so that the
spectrum of the invariant operator is known explicitly, then the
$\zeta$-function method is probably most convenient. %ref. 28
The generalized $\zeta$ function is defined by
\begin{equation}
\zeta(\nu)= \sum_{\bf n}(\mu^{-2}\lambda_{\bf n})^{-\nu},
\label{IRON8}
\end{equation}
where $\lambda_{\bf n}$ are the eigenvalues of the operator A on
the Euclideanized metric obtained by a Wick rotation to imaginary
time $\tau = it$.  Here a constant mass scale $\mu$ is introduced
to make the measure $d[\varphi]$ of the functional integral
dimensionless.
Using the regularization method of Dowker and Critchley, %ref. 28
one can express the one-loop effective potential as %ref. 14
\begin{equation}
V^{(1)}(\bar{\phi})= - \frac{1}{2} \hbar (Vol)^{-1}[\zeta'(0)+
\zeta (0)/\nu]. \label{IRON9}
\end{equation}
The ultraviolet divergence in $V^{(1)}$ can be canceled by the addition of
counterterms which is not our concern here. The regularized effective action is useful for the analysis of  the infrared behavior of quantum fields in a curved spacetime or spaces of nontrivial topology or finite extent, as shown below.
%\subsection{Symmetry behavior} \label{sub-IRSB}
%The symmetry of a quantum system in curved spacetime is determined by its infrared behavior while the dominant features near the symmetric state is largely determined by the {\bf zero mode} of the fluctuation operator which depends on the geometry and topology of the underlying space and requires a separate careful treatment.
Note, however that one-loop results are  insufficient. Leading infrared contributions in higher loop terms need be included by using composite operator techniques.

We learn from the work of \cite{HuOC86} that in spacetimes with some compact dimensions the lowest mode of the
fluctuation operator has the strongest effect on the symmetry
behavior of the system.
%This mode in fact dominates the infrared behavior: the system can be described by a field theory in a lower dimension.
When the lowest mode is massless it will give the dominant contribution to the effective
action.  The low energy behavior corresponds to a lower-dimensional
system. In what follows we first give a formal derivation of infrared dimensional reduction by examining
the result of the decoupling of the higher modes (or bands) in
the functional integral for the effective action.  We will then
give a physical explanation in terms of the correlation lengths
and the notion of effective IR dimension ({\bf EIRD}).  An alternative way of seeing this problem of
dimensional reduction is by spectral analysis.  This is applied
to direct product spaces with some compact dimensions and to
spaces which can be reduced to product spaces.

\subsection{Decoupling of the Higher Modes (or Bands)}
\label{sub-IRDC}

Let us examine systems where  the eigenvalues
of the fluctuation operator takes on a band structure.  By band
structure we mean that the eigenvalues occur in continua with each
continuum having a higher lowest eigenvalue than the previous one.
This is true for fields on spacetimes with compact sections or for
fields with discrete spectrum (e.g. the harmonic oscillator).  The
procedure is to expand the fields in terms of the band eigenfunctions
and convert the functional integral over the fields to an integral
over the amplitudes of the individual modes.

On a manifold with topology $R^d \times B^b$ where $B$ is compact,
consider quantum fields where the fluctuation operator $A$  in
(\ref{IRON4}) has the general form of a direct sum of operators
$D$ and $B$
\begin{equation}
A^{ab}(x,y) = D^{ab}(x) + B^{ab}(y)
\label{IRDC1}
\end{equation}
with coordinates $x$ on  $R^d$ and $y$ on $B^b$.  Assume that the
eigenvalues $\omega_n$  associated with the eigenfunctions
$\psi_n(y)$ of $B^{ab}$ are discrete:
\begin{equation}
B^{ab}\psi_n(y)= \omega^{ab}_{n} \psi _n(y)
\label{IRDC2}
\end{equation}
Decomposing the field $\varphi^a(x,y)$ in terms of $\psi_n(y)$
\begin{equation}
\varphi^a(x,y) = \sum_{n} \varphi^{a}_{n}(x)\psi_n(y)
\label{IRDC3}
\end{equation}
one obtains for the quadratic part of the action
\begin{equation}
\frac{1}{2} \int d^dxd^by\varphi^a A^{ab}\varphi^b = \frac{1}{2}
\int d^dx[\varphi^{a}_{n} f_{nm}D^{ab}\varphi^{b}_{m} +
\omega^{ab}_{n} f_{nm}
\varphi^{a}_{n} \varphi^{b}_{m}]    %equation (3.4)
\label{IRDC4}
\end{equation}
where $f_{nm} = \int d^by\psi_n(y)\psi_m(y).$  When $\varphi_n$
are properly normalized $f_{nm} = \delta_{nm}$ (we will make such
a choice here) the resulting theory in terms of the new fields
$\varphi^{a}_{n}$ will involve massive fields with masses
determined by the eigenvalue matrix $\lambda^{ab}_{in}$ of the
operators $\nabla^2+M_i$ [see (\ref{IRON7})], even if the fields
in terms of the old variables appeared massless.  We will take the
smallest eigenvalue to be given by $n=0$ and assume that its only
degeneracy is labeled by the indices $a$ and $b$. Assume also
that the operator $D^{ab}$ is simply minus the Laplacian $\nabla^2_d$
on $R^d$ times $\delta^{ab}$, the $n=0$ mode is then governed by
the action whose quadratic term is
\begin{equation}
\frac{1}{2} \int d^dx[\varphi^a_0(-\nabla^2_d)\delta^{ab}\varphi^b_0 +
\omega^{ab}_0 f_{00} \varphi^a_0\varphi^b_0] \label{IRDC5}
\end{equation}
Thus this appears like a $d$-dimensional field with an apparent
mass matrix  $\omega^{ab}_{0}$. For the case of an $N$ component
$\lambda \phi^4$ theory the action after this decomposition takes
the form
$$ S[\bar{\phi}+\varphi] = \int d^dx[\frac{1}{2}
\varphi^a_n(-\nabla^2_d \delta^{ab} + \omega^{ab}_n)\varphi^b_n$$
\begin{equation}
+ \frac{\lambda}{6} g^a_{n\ell m} \varphi^{b}_{\ell} \varphi^b_m +
\frac{\lambda}{4!} f_{kn\ell m}\varphi^a_k \varphi^a_n
\varphi^b_\ell
\varphi^b_m],   %equation (3.6)
\label{IRDC6}
\end{equation}
where $g^a_{n\ell m} = \int d^by \bar{\varphi}^a \psi_n \psi _\ell
\psi _m$ and $f_{kn\ell m} = \int d^by \psi_k \psi_n \psi _\ell
\psi _m.$  The effective action is now given by the functional
integral
\begin{equation}
e^{-\Gamma [\bar{\phi}]} = \int [d\varphi^a_n]e^{-S[\bar{\phi}
+ \varphi]}. \label{IRDC7}
\end{equation}
The interesting case occurs when the lowest eigenvalue approaches
zero.  At low energy the Appelquist-Carazzone decoupling theorem
assures us that with higher modes decoupled from the dynamics, the
infrared behavior is governed by the lowest band.  We are then
left with a purely lower dimensional theory.  The higher modes do
play a role in the ultraviolet divergences present in the theory
(e.g. renormalization problem in Kaluza-Klein theories) and
therefore determine the high energy running of coupling
constants, but the infrared region of the theory is governed by
the lower d-dimensional theory.

\subsection{Correlation Length and Effective Infrared Dimension}
\label{sub-IRED}

The above result of dimensional reduction from a formal
derivation of mode decoupling can be understood in a more
physical way by using the concept of effective infrared
dimensions (EIRD).  By effective IR dimension we mean the
dimension of space or spacetime wherein the system at low energy
effectively behaves.  One well-known example is the Kaluza-Klein
theory of unification and cosmology. There, starting with an
11-dimensional spacetime with full diffeomorphism symmetry, after
spontaneous compactification it reduces at energy below the
Planck scale to the physical four dimensional space with GL(4,R)
covariance and a seven-dimensional internal space with symmetry
group containing the standard $SU_3 \times SU_2 \times U_1$
subgroups of strong and electroweak interactions.  For observers
today of very low energy the effective IR dimension of spacetime
is four, even though the complete theory is eleven dimensional.
By the same token, Einstein's theory of general relativity can
presumably be regarded as the effective IR limit of an otherwise
more complete and fundamental theory of quantum gravity, such as
the superstring theory.  For curved-space symmetry breaking
considerations, the EIRD which the system ``feels'' is governed
by a parameter $\eta$ which is the ratio of the correlation
length $\Xi$ and the scale length $L$ of the background space
$\eta \equiv \Xi /L$.  For compact spaces like $S^4$, $L$ is
simply $2\pi$ times the ``radius'' of $S^4$, the only geometric
scale present.  For product spaces $R^d \times B^b$ with some
compact space $B$, there are two scale lengths: $L_b$ is finite in
the compact dimensions and $L_d = \infty$ in the non-compact
dimensions.  Examples are Kaluza-Klein theories d=4, b=1, 6, 7,
finite temperature field theory (imaginary time formalism,  $L =
\beta$ inverse temperature) d=3, b=1, and Einstein or the spatially-closed
Robertson-Walker universe d=1, b=3.\\

The symmetry behavior of the system (described here by a $\lambda
\phi^4$ scalar field as example) is determined by the {\em
correlation length} $\Xi$ defined as the inverse of the {\em
effective mass $ M_{eff}$} related to the effective potential
$V_{eff}$ by (we use subscript eff to denote quantities including
higher loop corrections)
\begin{equation}
\Xi^{-2} = \frac{\partial ^2V_{eff}}{\partial \bar{\phi}^2}
\mid_{\bar{\phi}_{min}} \equiv M^2_{eff}  =(^{curvature-induced \;
mass \; M^2_{1,2}}_{+ \; radiative \; corrections})    %equation (3.8)
\label{IRED1}
\end{equation}
It measures the curvature of the effective potential at a minimum
energy state $(\bar{\phi} = 0$ for the symmetric state, or the
false vacuum, $\bar{\phi}=\bar{\phi}_{min}$ for the
broken-symmetry state or the true vacuum.)  The effective mass is
defined to include radiative corrections to the same order
corresponding to the effective potential.  (This quantity is
called the generalized susceptibility function in condensed
matter physics.)  The critical point of a system is reached when
$\Xi \rightarrow \infty$ or $ M_{eff} \rightarrow 0.$  In flat or
open spaces or for bulk systems, the critical point can be
reached without restriction from the geometry (note that in
dynamical situations, exponential expansion can effectively
introduce a finite size effect equivalent to event horizons, see
\cite{HuEdmonton}).  However, in spaces with compact dimensions,
the correlation length of fluctuations can only extend to
infinity in the remaining non-compact dimensions, and thus the
critical behavior becomes effectively equivalent to a lower
d-dimensional system.  One can also think of $\Xi$ as the Compton
wavelength $\Lambda = 2\pi/M_{eff}$ of a system of
quasi-particles with effective mass $M_{eff}$.  Any fine
structure of the background spacetime with scale $L$ is relevant
only if $\Lambda \leq L$.  Thus when $\Lambda$ is small or $\eta
\ll 1$, (far away from critical point, at higher energy, with
higher mode contributions) it sees the details of a spacetime of
full dimensionality.  At this wavelength, the apparent size of the
universe is large in both compact and non-compact dimensions.
When $\Lambda \rightarrow \infty$ or $\eta \gg 1$ (near critical
point, IR limit, lowest mode dominant) finer features in the compact dimensions will not be important.  The apparent
`size' of the universe will be measured by the non-compact
directions and the EIRD is measured by the number of non-compact
dimensions.  The value of $\eta$ getting very large is an
indication of when dimensional
reduction can take place.\\

Notice that in flat space $(R^d)$ critical phenomena the effective
potential $V_{eff}$ (free energy density) depends on the coupling
constants of fields which run with energy and temperature.  In
curved-space coupling parameters run also with curvature or the
scale length of the space.  This makes the concept of EIRD even
more interesting, as there is now an interplay between $\Xi$ and
$L$, and $\eta$ can either decrease or increase with curvature.
For example, for $\lambda \phi^4$ fields in the Einstein universe
near the symmetric state $\bar{\phi}=0$, the EIRD is equal to 1
but near the global minimum of broken symmetry state
$\bar{\phi}=\bar{\phi}_{min}$ is equal to 4. Near the symmetric
state, $\eta \gg 1$ signifies reduction of EIRD to one. This is
consistent with the theorem of Hohenberg, Mermin and Wagner (for
statistical mechanics on a lattice) and Coleman (for continuum
field theory) which states that in dimensions less than or equal
to two, the infrared divergence of the scalar field is so severe
that there could be no possibility of spontaneous symmetry
breaking: the only vacuum expectation value for $\bar{\phi}$
allowed is zero.  Away from the region $\bar{\phi} \simeq 0$ the
one-dimensional behavior no longer prevails.  Indeed a global
minimum of the effective potential exists at $\bar{\phi}_{min}$.
Near $\bar{\phi}_{min}$, $\eta \ll 1$ and decreases with curvature. Thus the apparent size
of the universe near the global minimum actually increases with
increasing curvature.  There is therefore no dimensional
reduction and the system has a full 4-dimensional IR behavior.  A
transition  to the asymmetric ground state is not precluded as
symmetry breaking via tunneling is in principle possible.  The
complete picture extending from $\bar{\phi}=0$ to
$\bar{\phi}=\bar{\phi}_{min}$ is a combination of one dimensional
and four dimensional infrared behavior.  Similar arguments can be
applied to other spacetimes or field theories.  Using this notion
one can understand, for example, why it is often said that at high
temperatures (small radius limit of $S^1$) the finite temperature
theory becomes an effective three-dimensional theory.

\subsection{Dimensional Reduction:  An Eigenvalue Analysis}
\label{sub-IREG}

In the above we have introduced the notion of effective infrared
-dimension and suggested the ratio of the correlation length $\Xi$ to
the geometric scale of spacetime $L$ as a measure of the conditions for
the system to behave effectively as in a reduced dimension in the
infrared regime.  We suggested that for product spaces $R^d \times
B^b$ with some noncompact dimension d, the effective IR dimension is
usually just $d$.  We will now verify this assertion by
analyzing the spectrum of the fluctuation operators in these
spacetimes directly.        \\
%We will first discuss direct product spaces and then discuss spacetimes which can be reduced to product spaces in some limit of continuous deformation.\\

Consider for simplicity {\em direct product spaces}.  Examples are a) physical cosmological spacetimes
with  topology $R^1(time) \times S^3(or \; R^3, H^3, T^3)$, b) Kaluza-Klein cosmology with $M^4$ (Minkowski) $\times S^7$ (or other
internal space), c) finite-temperature (imaginary-time) theory has $R^3 \times S^1$.  Take a simple example $S^2 \times S^1$ for illustration.
(This could be the spatial geometry of a ``handled'' Gowdy universe.)
Similar reasoning can be extended to a wide range of product
spaces.\\

From (\ref{IRON7}) the wave operator $A$ governing the fluctuation
fields in the large $N$ limit is $A \equiv \nabla^2 + M^2_2$, where
$\nabla^2$ is the Laplace-Beltrami operator on a general curved
spacetime. For $S^2 \times S^1$ with radii $a_2$ and $a_1$
respectively,  %$\Delta$ (for a three dimensional space we replace by $\Delta$)
$\nabla^2$   is a sum of the total angular momentum
operator $L$ on $S^2$, and $L_z$ on $S^1$
\begin{equation}
\nabla^2 = \frac{L^2}{a^2_2} + \frac{L^2_z}{a^2_1} \label{IREG1}
\end{equation}
The eigenfunction is a product of $Y_{\ell m}(\theta,\phi)e^{in\chi}$
belonging to the eigenvalues
\begin{equation}
\kappa^2_{\bf n} = \frac{\ell(\ell + 1)}{a^2_2} +
\frac{n^2}{a^2_1} \; , \label{IREG2}
\end{equation}
where ${\bf n}= (\ell , m,n), \ell = 0 \cdot \cdot \infty, m =
-\ell$ to $\ell, n = 0 \pm 1, \cdot \cdot \cdot$ The eigenvalues
of $A$ are then $\lambda_{\bf n} = \kappa^2_{\bf n} + M^2_2$.  In
the infrared region, we are interested in the contribution of the
lowest eigenvalue (zero mode) to the effective potential.  We will
consider the two limiting cases of 1)  $S^2 \times R^1$ and 2)
$R^2 \times S^1$ obtained when $a_1$ and $a_2 \rightarrow \infty$
respectively, and show that the EIRD is
equal to 1 and 2 respectively.\\

The effective potential $V_{eff}$ can be constructed from the
$\zeta$ function $\zeta(0)$ and its derivative $\zeta '(0)$ by
(\ref{IRON8}).
\begin{equation}
\zeta(\nu) = \sum_{\bf n} (\mu^{-2}\lambda_{\bf n})^{-\nu} =
\mu^{2\nu} \sum_{\ell,n} (2\ell +1) [\frac{\ell (\ell +
1)}{a^2_2} +
\frac{n^2}{a^2_1} + M^2_2]^{-\nu}  %Equation(3.13)
\label{IREG3}
\end{equation}
where $2\ell +1$ is the degeneracy of $m$. In the limiting cases
the summation over discrete quantum numbers will be replaced by
integrals of the form $\int k^{D-1}dk$.  We will derive the
dimensionality of the reduced system by finding $D$.  Thus in case
a) \ $a_1 \rightarrow \infty$, the lowest eigenvalues belong to
the band $\ell =0$  as $a_1 \rightarrow \infty,\;  k = n/a_1$
assumes continuous value and $\zeta (\nu)$ becomes
\begin{equation}
\zeta(\nu) \sim a_1 \int^{\infty}_{- \infty} dk(k^2 +
M^2_2)^{-\nu}, \; a_1 \rightarrow \infty, ~~ M_2a_2 << 1
\label{IREG4}
\end{equation}
This is a one-dimensional integral.  In case (b) \  $a_2
\rightarrow \infty$, the lowest eigenvalues come from the lowest
band n=0 given by the first term in (\ref{IREG2}), where $\ell$
assumes continuous values as $a_2 \rightarrow \infty$.  Now call
$k = \ell/a_2, \zeta(\nu)$ becomes ($n=0$)
\begin{equation}
\zeta(\nu) \sim 2a_2 \int^{\infty}_{o} dk k(k^2 + M^2_2)^{-\nu}, \; a_2
\rightarrow \infty \; ,\; M_2a_1 << 1   %equation (3.15)
\label{IREG5}
\end{equation}
The extra factor of $k$ comes from the degeneracy of $m$ for
nonzero
$\ell$.  The integral is thus 2-dimensional, as expected.

% $S^3 \rightarrow S^2 \times S^1.$

Physically, the two limiting cases may represent two different
symmetry states of the system: one with cylindrical symmetry where
the quantum states with good quantum number m are eigenstates of
the $L_z$ operator, the other with spherical symmetry with good
quantum numbers $\ell$ associated with the $L^2$ operator (on
$S^2$).  As a result of symmetry breaking the original system can
end up in one of these states with a different symmetry.  External
perturbations with a particular symmetry and sufficient magnitude
will influence the selection of the end state of the system.  A
simple example in elementary quantum mechanics is the Zeeman
versus the Paschen-Bach effects associated with many electron
atoms subjected to an external magnetic field.  The total
magnetic quantum number will become a good quantum number at
very  strong fields.  More sophisticated examples can be found in
the consideration of classes of internal spaces admitting larger
symmetry groups with the strong and electroweak gauge groups as
subgroup in the Kaluza-Klein theory.  By the same analysis, it is
not difficult to see that the EIRD of any homogeneous cosmology
with compact spatial section is one, the spatial metrics of the
Einstein-Rosen waves $S^2 \times R^1$, the Kantowski-Sachs
universe with spatial metric $d\ell^2 = dz^2 + a^2d\Omega^2_2$,
black hole spacetimes and the whole class of stationary
axisymmetric metrics in relativistic astrophysics have EIRD=2.
Similarly, the (imaginary time) finite temperature theory $R^3
\times S^1$ has EIRD three and the Kaluza-Klein theory on $M^4
\times B^b$ with compact internal space $B$ has EIRD four.

It is important to recognize that this infrared dimensional reduction feature is not due directly to the topology of the underlying space, but rather the `finite size' of the system. Spaces which are not direct product spaces but which can approach product spaces in the limit of extreme deformations or large perturbations have similar IR behavior. In the example of direct product spaces shown above, $S^2 \times S^1$ can be shown to be the extreme deformation limit of $S^3$:  The eigenfunctions of  $S^3$ which are the  hyperspherical harmonics   $Y_{n\ell m}(\chi, \theta, \phi)$
can be expressed as a product of the spherical harmonics  $Y_{n \ell m}(\theta, \phi)$  and the Gegenbauer polynomial $G_{n\ell}(\chi)$.  The eigenfunctions  $e^{in\chi}$ on $S^1$ we saw above are  obtained as limits of  $G_n(\chi)$ as the angular momentum of $S^2$ is  decoupled from the``radial" equations governing $G_{n\ell}$. Another example given
in \cite{HuOC86} is the (static) Taub universe which is a deformed Einstein Universe.   An  eigenvalue analysis there provides a vivid illustration of this infrared dimensional reduction effect. To the extent that the large scale features of a quantum system shows up near criticality this reveals the surprising effect of its underlying spacetime structure.

% \subsection{Quantum Criticality and Entanglement in $O(N)$ model} Sachdev's paper the notion of small and large
% \subsection{UV/IR duality from string theory, AdS/CFT correspondence} UV/IR further mixes up

%\newpage

\section{Summary} %: What really is macroscopic? Is MQP really so strange?} \subsection{What is MQP?}
%determines which level of structure we focus on. Born-Oppenheimer approx for molecules ignore nucleus in the atom. Our CHY paper also touched on this point.)

In our first paper we explored macroscopic quantum phenomena (MQP) from a large $N$ perspective, where $N$ is the number of particles or components in a many-body quantum system.  In this paper we explore MQP from the perspective of correlations and interactions which exist between components of the quantum system, how the coupling strength varies (runs) with energy or scale, especially how its infrared behavior at the critical point determines or affects the macroscopic features of a  quantum system. We began with a review of the way how $N$, the number of particles in a quantum system, enters in the thermodynamic functions in quantum statistical mechanics. The (grand) partition function of a system at finite temperature (and chemical potential) in the (grand) canonical ensemble contains an explicit dependence of $N$, with proper factors accounting for whether the particles are distinguishable or identical.  We note that the meaning of quantum in QSM refers specifically to spin-statistics (boson or fermions) and distinguishability,  whereby classical and quantum are clearly divided. This restricted sense of quantumness in QSM is different from the other meanings or criteria of `quantumness' which define MQP, i.e., quantum coherence, fluctuations, correlation and entanglement associated with the full many-body  wave function of a quantum system.

We then explore the issue of how MQP can be measured or understood from the behavior of quantum correlations which exist in a quantum system and how the interaction strengths change with energy or scale, under ordinary situations and when the system is near its critical point. Noting that quantum correlation is further related to quantum fluctuations and quantum coherence, we used three examples from known results to illustrate these aspects: a) the existence of an H-theorem at the next-to-leading order large $N$ expansion for an $O(N)$ quantum mechanical model, with entropy generation signifying the  emergence of macroscopic (thermodynamic) properties. b) the nonequilibrium dynamics of $N$ atoms in an optical lattice under the large $\cal N$ (field components), $2PI$ and second order perturbative expansions, illustrating how $N$ and $\cal N$ enter in these  three aspects of quantum correlations, fluctuations and coupling strength. c) the infrared behavior of an interacting quantum system, we discuss the conditions where dimensional reduction shows up. The effective IR dimension is determined by the spectrum of the Laplacian of an interacting quantum system, in particular, whether a lowest eigenmode exists.  On the same theme,  we also discuss how the effective field theory concept bears on MQP: the running of the coupling parameters with energy or scale imparts a dynamical-dependent  and an interaction-sensitive definition of `macroscopia'.

Quantum entanglement will be the theme of our third essay where we intend to focus on how this uniquely quantum feature shows up in quantum many-body systems and inquire to what extent it quantifies a quantum system as micro, meso versus macro. In this enquiry we will necessarily also touch on related subjects such as `small systems' and emergent concepts in the new field known as quantum thermodynamics (ostensibly this is not the thermodynamics of quantum systems, treated in QSM textbooks, where quantum pertains to spin-statistics of particles). We can see why this is relevant to MQP in a simplified way: thermodynamics is a macroscopic theory valid when the volume V and number of constituents $N$ both go to infinity while their ratio $N$/V remains constant. Quantum dynamics is used for the description of micro-objects or small systems. As one increases the `size' (number and volume, say) of a quantum system, in what parameter ranges will a thermodynamic description of such a many body quantum system become useful, at least begin to make sense? These inquires into the meaning of MQP will unavoidably lead us to confront some foundational issues of quantum and statistical mechanics.

\ack

We thank Esteban Calzetta for discussions at the initial stage of this work. BLH enjoyed the warm hospitality of the National Center for Theoretical Sciences (South) and the Department of Physics of National Cheng-Kung University, Tainan, Taiwan while part of this work was done. This work is supported by the National Science Council of Taiwan under grant NSC 97-2112-M-006-004-MY3, the US NSF grants PHY-0801368 to the University of Maryland. Two of the authors spent some time at the KITPC program on Quantum Open Systems supported by the Project of Knowledge Innovation Program (PKIP) of Chinese Academy of Sciences, Grant No. KJCX2.YW.W10.\\


\begin{thebibliography}{99}

\bibitem{MQP1} Chou C H, Hu B L and Suba\c{s}\i Y 2011 {\it Preprint} quant-ph/1106.0556v1

\bibitem{HuangSM} Huang K 1987 {\it Statistical Mechanics} 2nd ed. (Wiley)

\bibitem{ZurekPT} Zurek W H 1991 {\it Phys. Today} {\bf 44} 36

\bibitem{QdecBook} Guilini D et al 2003 {\it Decoherence and the Appearance of a Classical World in Quantum Theory} 2nd ed. (Berlin: Springer-Verlag)

\bibitem{Pathria} Pathria R K and Beale P D 1996 {\it Statistical Mechanics} 2nd ed. (Butterworth-Heinemann)

\bibitem{CHH} Calzetta E, Ho Kwan-yuet and Hu B L 2010 Vortex Formation in Two-Dimensional Bose Gas, {\it J. Phys.} {\bf B}: At. Mol. Opt. Phys. {\bf 43}, 095004.

\bibitem{E/QG} Hu B L 2009 {\it J. Phys. Conf. Ser.} {\bf 174} 012015 {\it Preprint} gr-qc/09030878

\bibitem{RG2000} Calzetta E, Hu B L and Mazzitelli F D 2001 {\it Physics Report} {\bf 352} 459-520 {\it Preprint} hep-th/0102199

\bibitem{SILARG} Hu B L 1991 Coarse-Graining and Backreaction in Inflationary and Minisuperspace Cosmology {\it Invited Lectures at the Seventh International Latin-American Symposium on General Relativity (SILARG VII)} Proceeding appeared as {\it Relativity and Gravitation: Classical and Quantum} J. D'Olivio et al (eds) (Singapore: World Scientific)

\bibitem{QUPON} Hu B L 2005 The Universe as an Ultimate Macroscopic Quantum Phenomenon, {\it Invited talk at the Quantum Physics of Nature (QUPON) Conference} Vienna, Austria

\bibitem{BHmicro} Jacobson T A 1995 Introduction to Black Hole Microscopy {\it Utrecht} THU-95/23, {\it Preprint} hep-th/9510026

\bibitem{CHcorhis} Calzetta E and Hu B L 1993 Decoherence of correlation histories {\it Directions in General Relativity} vol. 2 Hu B L and Jacobson T A (eds) (Cambridge: Cambridge Univ. Press) {\it Preprint} hep-th/9501040

\bibitem{Cirac} Verstraete F, Popp M and Cirac J I 2004 {\it Phys. Rev. Lett.} {\bf 92} 027901

\bibitem{mea} Calzetta E and Hu B L 1999 {\it Phys. Rev.} D {\bf 61} 025012

\bibitem{CH95} Calzetta E and Hu B L 1995 Correlations, decoherence, dissipation and noise in quantum field theory {\it Heat Kernel Techniques and Quantum Gravity} vol. 4 of {\it Discourses in Mathematics and Its Applications} ed Fulling S A (A\&M Univ. Press, College Station, TX) {\it Preprint} hep-th/9501040

\bibitem{CH00} Calzetta E and Hu B L 2000 {\it Phys. Rev.} D {\bf 61} 025012

\bibitem{CH03} Calzetta E A and Hu B L 2003 {\it Phys. Rev.} D {\bf 68} 065027

\bibitem{MACDH00} Mihaila B, Athan T, Cooper F, Dawson J and Habib S 2000 {\it Phys. Rev.} D {\bf 62} 125015

\bibitem{MDC01} Mihaila B, Dawson J and Cooper F 2001 {\it Phys. Rev.} D {\bf 63} 096003

\bibitem{balescu} Balescu R 1975 {\it Equilibrium and Nonequilibrium Statistical Mechanics} (New York: John Wiley \& Sons)
Balescu R 1997 {\it Statistical Dynamics, Matter out of Equilibrium} (London: Imperial College Press)

\bibitem{KB62} Baym G and Kadanoff L 1962 {\it Quantum Statistical Mechanics} (New York: Benjamin)

\bibitem{CH88} Calzetta E and Hu B L 1988 {\it Phys. Rev.} D {\bf 37} 2878

\bibitem{ctp} Schwinger J 1960 {\it Field Theory Methods in non-field-theory contexts} Brandeis Summer Institute in Theoretical Physics.
Keldysh L V 1964 {\it Zh. Eksp. Teor. Fiz.} {\bf 47} 1515 (1965 {\it Sov. Phys. JETP} {\bf 20} 1018)

\bibitem{2pi} Cornwall J M, Jackiw R and Tomboulis E 1974 {\it Phys. Rev.} D {\bf 10} 2428

\bibitem{HuPhysica} Hu B L 1989 {\it Physica} A {\bf 158} 399

\bibitem{HuBanff} Hu B L 1994 Quantum statistical fields in gravitation and cosmology {\it Proc. Third
International Workshop on Thermal Field Theory and Applications} Kobes R and Kunstatter (eds) (Singapore: World Scientific) {\it Preprint} gr-qc/9403061

\bibitem{fdr} Hu B L and Sinha S 1995 {\it Phys. Rev.} D {\bf 51} 1587

\bibitem{KacLog} Kac M and Logan J 1979 Fluctuations {\it Fluctuation Phenomena} eds Montroll E and Lebowitz J (New York: Elsevier)

\bibitem{Spohn} Spohn H 1991 {\it Large Scale Dynamics of Interacting Particles} (Berlin: Springer-Verlag)

\bibitem{Hab90} Habib S 1990 {\it Phys. Rev.} D {\bf 42} 2566

\bibitem{AABBS02} Aarts G, Ahrensmeier D, Baier R, Berges J and Serreau J 2002 {\it Phys. Rev.} D {\bf 66} 045008

\bibitem{HuPav} Hu B L and Pavon D 1986 {\it Phys. Lett.} B {\bf 180} 329

\bibitem{HKMP96} Habib S, Kluger Y, Mottola E and Paz J 1996 {\it Phys. Rev. Lett.} {\bf 76} 4660

\bibitem{Habib08} Habib S 2004 Gaussian dynamics is classical dynamics {\it Preprint} quant-ph/0406011

\bibitem{CH08} Calzetta E and Hu B L 2008 {\it Nonequilibrium Quantum Field Theory} (Cambridge: Cambridge University Press)

\bibitem{Fishers} Fisher M P A, Weichman P B, Grinstein G and Fisher D S 1989 {\it Phys. Rev.} B {\bf 40} 546

Oosten D van, Straten P van der and Stoof H T C 2001 {\it Phys. Rev.} A {\bf 63} 053601

\bibitem{Rey04}Rey A M, Hu B L, Calzetta E, Roura A and Clark C W 2004 {\it Phys. Rev.} A {\bf 69} 033610

\bibitem{Morgan} Morgan S A 2000 {\it J. Phys} B {\bf 33} 3847

\bibitem{BurnettComparisons} Hutchinson D A W {\it et al.} 2000 {\it J. Phys} B {\bf 57} 3825

\bibitem{Martin} Hohenberg P and Martin P 1965 {\it Ann. Phys. (N.Y.)} {\bf 34} 291

\bibitem{Griffin} Shi H and Griffin A 1998 {\it Phys. Rep} {\bf 304} 1

Griffin A 1996 {\it Phys. Rev.} B {\bf 53} 9341

\bibitem{Giorgini} Giorgini S 1998 {\it Phys. Rev.} A {\bf 57} 2949

Giorgini S 2000 {\it Phys. Rev.} A {\bf 61} 063615
%
%\bibitem{Liboff03} Liboff R L 2003 {\it Kinetic Theory} 2nd ed (New York: Springer-Verlag)

\bibitem{Burnett} Proukakis N P and Burnett K 1995 {\it J. Res. Natl. Inst. Stand. Technol.} {\bf 101} 457

Proukakis N P, Burnett K and Stoof H T C 1998 {\it Phys. Rev.} A {\bf 57} 1230

Kohler T and Burnett K 2002 {\it Phys. Rev.} A {\bf 65} 033601

%\bibitem{CHR} Calzetta E, Hu B L and Rey A M 2006 {\it Phys. Rev.} A {\bf 73} 023610

\bibitem{CHR} Calzetta E, Hu B L and Ramsey S A 2000 {\it Phys. Rev.} D {\bf 61} 125013

\bibitem{Walls} Wright E M, Wong T, Collett M J, Tan S M and Walls D F 1997 {\it Phys. Rev.} A {\bf 56} 591

\bibitem{Smerzi} Smerzi A and Raghavan S 2000 {\it Phys. Rev.} A {\bf 61} 063601

\bibitem{HuOC86} Hu B L and O'Conner D J 1987 {\it Phys. Rev. D} {\bf 36} 1701

\bibitem{HuEdmonton} Hu B L 1987 Dynamical Finite Size Effect, Inflationary Cosmology and Thermal Particle Production, invited talk given at the CAP-NSERC Summer Institute in Theoretical Physics, Edmonton, Canada. Proceedings edited by K. Khanna, G. Kunstatter and H. Umezawa (Singapore: World Scientific Publishing Co. 1988).

%
%\bibitem{SuFu90} Sumino M and Fukuda R 1990 {\it Journal of The Physical Society of Japan} {\bf 59} pp 3553-63
%
%\bibitem{CRV} Calzetta E, Roura A and Verdaguer E 2003 {\it Physica} A {\bf 319} 188
%
%\bibitem{Habib08} Habib S 2004 Gaussian dynamics is classical dynamics {\it?Preprint} quant-ph/0406011
%
%\bibitem{Anastopoulos10} Anastopoulos C, Kechribaris S and Mylonas D 2010 {\it Phys. Rev.} A {\bf 82} 042119
%
\end{thebibliography}
\end{document}